\documentclass[proceedings]{JHEP37}
\usepackage{eurosym}
\usepackage{amssymb}
\usepackage{amsfonts,cite}
\usepackage{amsmath}
\usepackage{graphicx}
\usepackage{multirow}
\usepackage{epsfig}
\usepackage[utf8]{inputenc}

\DeclareMathOperator\arcsec{arcsec}

\newbox\mybox

\newcommand\fverb{\setbox\mybox=\hbox\bgroup\verb}
\newcommand\fverbdo{\egroup\medskip\noindent\fbox{\unhbox\mybox}\ }
\newcommand\fverbit{\egroup\item[\fbox{\unhbox\mybox}]}
\conference{Infinite Weyl groups with their associated Calogero models}
\abstract{We propose generalizations of Calogero models that exhibit invariance with respect to the infinite Weyl groups of affine, hyperbolic, and Lorentzian types. Our approach involves deriving closed analytic formulas for the action of the associated Coxeter elements of infinite order acting on arbitrary roots within their respective root spaces. These formulas are then utilized in formulating the new type of Calogero models.}

\title{Infinite affine, hyperbolic and Lorentzian Weyl groups with their associated Calogero models}
\author{Francisco Correa$^\circ$, Andreas Fring$^\bullet$ and Octavio Quintana$^\star$ \\
 $\circ$ Instituto de Ciencias F{\'{\i}}sicas y Matem{\'{a}}ticas,
 Universidad Austral de Chile, \\
 $\,\,$ Casilla 567, Valdivia, Chile\\
 $\bullet$ Department of Mathematics, City, University of London,\\
 $\,\,$ Northampton Square, London EC1V 0HB, UK \\
 $\star$ Facultad de Física, Pontificia Universidad Cat\'olica de Chile, Vicu\~na Mackenna 4860, \\ 
 $\,\,$ Santiago, Chile
 
$\,\,$ francisco.correa@uach.cl, a.fring@city.ac.uk, oaquintana@uc.cl}

\input{tcilatex}
\begin{document}

\section{Introduction}
Calogero models \cite{Cal1,Cal2,Cal3}, along with their closely related generalisations to Calogero-Moser-Sutherland models \cite{Mo,Suth3,Suth4} and their $\cal{PT}$-symmetric extensions \cite{Mon1} are known to be invariant under the Weyl group related to finite semi-simple Lie algebras. This symmetry property, combined with their Liouville integrability, is largely responsible for the exact solvability of these systems in the classical as well as in their quantum theoretical versions. One may for instance utilise directly the Weyl invariant polynomials as a starting point to set up exactly solvable models \cite{HR1,HR3,Fring:2004zw} or in turn exploit the Weyl symmetry in the construction of differential-difference (Dunkl) operators \cite{dunkl89,carrillo21} that can be used in the construction of the orthogonal eigenstates in the quantum theory.  

Here, we propose generalisations of these models to make them invariant under the action of Weyl groups of infinite order such as the affine, hyperbolic and Lorentzian generalisations of the finite groups. Thus, we are seeking generalisations of the standard Calogero Hamiltonians 
\begin{equation}
H = \frac{1}{2} p^2 + \sum_{\alpha \in \Delta_{\bf{g}} }  \frac{c_{\alpha}}{\left( \alpha \cdot q \right)^2 } = \frac{1}{2} p^2 +  \sum_{i=1}^r  \sum_{n=1}^h  \frac{ c_{in}}{\left[ \sigma^n ( \gamma_i) \cdot q \right]^2 } ,\label{stanCalo1}
\end{equation}
where $\alpha$ are the roots in the root space $\Delta_{\bf{g}} \in \mathbb{R}^\ell $ associated to the finite Lie algebra $\bf{g}$ of rank $r$,  $c_{\alpha}$ and $ c_{in}$ are some real coupling constants, $\sigma$ is a Coxeter element, $\gamma_i$ is a representative of the $i$th orbit and $q=(q_1,\ldots, q_\ell)$, $p=(p_1,\ldots, p_\ell)$ are the $\ell$ coordinates and momenta of the system, respectively. Since the sum in the potential extends over the entire root space, the invariance of the Hamiltonian under the Weyl group associated to $\bf{g}$ is built-in by construction.

  Both variants of the Hamiltonian in (\ref{stanCalo1}) are equivalent. The first version is more formal as it still involves the task of identifying the entire root system of ${\bf{g}}$. This is made explicit in the second formulation in which the entire root space of the $r \times h$ roots is systematically generated from summing over $r$ orbits generated by the Coxeter element $\sigma$ of order $h$ action on some well-known representatives $\gamma_i$. In the finite dimensional case these representatives can be chosen to be the simple roots dressed by plus-minus signs assigned according to a bicoloured Dynkin diagram \cite{carter1989simple,dorey1991root,Mass2}. Thus, the challenge to generalise the symmetry properties of these models to infinite Weyl groups consists in the first version to listing systematically the entire infinite root spaces of the respective groups or in the second version to finding closed formulae for infinite consecutive actions of the Coxeter element on arbitrary roots together with the identification of the representatives $\gamma_i$ of the respective orbits. As we shall discuss below, the first version involved as many infinite sums as the rank of the algebra over integers constrained by a Diophantine equation. The second formulation only consists of a possibly finite sum over the orbit representatives and only one infinite sum over powers of the Coxeter element.  
  
  So far only one attempt has been made to formulate such theories, namely for the special case of the hyperbolic $AE_3$-algebra \cite{lechtenfeld2022}, where the left variant of the Hamiltonian in (\ref{stanCalo1}) was explored for that particular symmetry. Here we will explore both version and also elaborate on their correspondence. In the first instance we will extend the action of the repeated action of Coxeter elements to infinite order and derive closed analytical formulas for some specific algebra, which can not be found in the literature. Subsequently we attempt to carry out the sum over the entire orbits of these Coxeter elements to find simpler versions of the invariant potentials. For Toda field theories, that require only simple roots of the respective root spaces in their formulation, generalisations to Lorentzian and hyperbolic types have recently been explored in \cite{fring2021lorentzian}. Here we will largely adopt the notation from there and \cite{gaberdiel2002class,fring2019n,ThesisSam}.  In that context, the study of models based on extended Kac-Moody algebras is partly motivated by the success of the systematic description of conformal and massive Toda field theories in terms of finite dimensional and affine Kac-Moody algebras, respectively. The extended versions of hyperbolic and Lorentzian type algebra appear to provide a natural framework for a large class of nonintegrable theories. The particular extensions to $E_{10}$ and $E_{11}$ ($(E_8)_{-1}$ and $(E_8)_{-2}$ in our notation) play a prominent role in string/M-theory, see e.g. \cite{west2000hidden,west2001e11,damour200210,bossard21}. The mathematics developed here may also turn out to be useful in that context.
  
  Our manuscript is organised as follows: In section 2 we discuss in detail the infinite dimensional Weyl group associated to the $\mathbf{(A}_{2}\mathbf{)}_{-2}$-Lorentzian Kac-Moody algebra, together with their hyperbolic, affine and finite subalgebras. In particular we systematically derive closed analytic formulas for the action of any power of the Coxeter elements of infinite order on arbitrary roots. In section 3 we discuss how invariants for these type of infinite Weyl groups can be constructed by exploiting the possibility of the bicolouration, exemplified for the $\mathbf{(A}_{2}\mathbf{)}_{-3}$-Lorentzian Kac-Moody algebra. We prove that the Kostant identity also holds for these algebras, but integer exponents do not exist, although their analogues still add up pairwise to $\pi$. In section 4 we use the properties of the infinite Weyl groups to propose generalised versions of Calogero models invariant under Weyl groups of infinite order of affine, hyperbolic and Lorentzian type. Our conclusions are stated in section 5.

\section{Weyl groups of the $\mathbf{(A}_{2}\mathbf{)}_{-2}$-Lorentzian Kac-Moody algebra}
The essential tool in the study and classification of Lie algebras is the Cartan matrix $K$. It encodes the structure of the algebras by capturing the information about their weight and root systems, see e.g. \cite{Bou,hum2012}. In turn, this structures are inscribed into the corresponding Dynkin diagrams. The Cartan matrix for the $\mathbf{(A}_{2}\mathbf{)}_{-2}$-Lorentzian Kac-Moody algebra and the corresponding Dynkin diagram drawn in the usual conventions are\\
\setlength{\unitlength}{0.58cm} 
\begin{picture}(13.00,5.)(0.0,2.5)
	\thicklines
	\put(1.5,5.0){\Large{ $\mathbf{(A}_{2}\mathbf{)}_{-2}:$ }}
	\put(5.3,6){{$\small{\alpha_{1}}$}}
	\put(5.3,4){{$\small{\alpha_{2}}$}}
	\put(6.2,4.2){\line(0,1){2}}
	\put(6.0,6){\Large{$\bullet$}}
	\put(6.0,4){\Large{$\bullet$}}
	\put(7.2,5.2){\line(-1,1){0.9}}
	\put(7.2,5.2){\line(-1,-1){0.9}}
	\put(7.0,5){\Large{$\bullet$}}
	\put(7.4,5.2){\line(1,0){0.9}}
	\put(6.8,4.5){{$\small{\alpha_{0}}$}}
	\put(8.0,5){\Large{$\bullet$}}
	\put(8.2,5.2){\line(1,0){0.9}}
	\put(7.8,4.5){{$\small{\alpha_{-1}}$}}
	\put(9.0,5){\Large{$\bullet$}}
	\put(9.0,4.5){{$\small{\alpha_{-2}}$}}
	\put(13.0,5){$\!\!\! K_{ij}=2 \frac{\alpha_i \cdot \alpha_j}{\alpha_j \cdot \alpha_j}=\left(
		\begin{array}{ccccc}
			2 & -1 & 0 & 0 & 0 \\
			-1 & 2 & -1 & 0 & 0 \\
			0 & -1 & 2 & -1 & -1 \\
			0 & 0 & -1 & 2 & -1 \\
			0 & 0 & -1 & -1 & 2 \\
		\end{array}
		\right)_{ij},$}
\end{picture}\\
where we label rows and columns in the order $-2,-1,0,1,2$. Being of Lorentzian type we expect $K$ to be nonsingular and to possess precisely one negative eigenvalue, see \cite{gaberdiel2002class}. Indeed, the eigenvalues are
\begin{equation}
	1, \quad 3, \quad 2 + \frac{2}{\sqrt{3}} \cos \lambda \pm 2 \sin \lambda, \quad 2 - \frac{4}{\sqrt{3}} \cos \lambda<0, 
	\qquad \lambda = \frac{1}{3} \arctan \left(\sqrt{ \frac{37}{3} } /3  \right).
\end{equation}
In order to reproduce $K$ from the product of roots $\alpha_i$ as specified, we need to re-define the standard Euclidean inner product in the $\Delta_{\mathbf{(A}_{2}\mathbf{)}_{-2}}$-root space. Here we choose a 3+4 dimensional representation for the  $\mathbf{(A}_{2}\mathbf{)}_{-2}$-roots $\alpha_i$, $i=-2,\cdots,2$, in which the positive definite part is taken in the standard three dimensional representation, see e.g. \cite{Bou,hum2012}, and the extensions are constructed following the prescription in \cite{gaberdiel2002class,fring2019n} by adding two copies of the self-dual Lorentzian lattice $\Pi^{1,1}$-lattice
\begin{eqnarray}
	\alpha _{1} &=&\left( 1,-1,0;0,0,0,0 \right)
	,\quad \alpha _{2}=\left(0,1,-1;0,0,0,0 \right) , \\
	\alpha _{0}&=&\left(-1, 0, 1; 1, 0, 0, 0 \right) , \quad
	\alpha _{-1} =\left( 0,0,0;-1,1,0,0 \right) ,\quad \alpha _{-2}=\left(
	0,0,0;1,0,-1,1\right) . \notag
\end{eqnarray}
The Lorentzian inner product for two 3+4 dimensional vectors $x=(x_1, \ldots x_7)$, $y=(y_1, \ldots y_7)$ is then taken to be
\begin{equation}
x\cdot y = x_1 y_1 + x_2 y_2 + x_3 y_3 - x_4 y_5 - x_5 y_4- x_6 y_7 - x_7 y_6. \label{definner}
\end{equation}
Demanding the length of an arbitrary root 
\begin{equation}
	\alpha = p \alpha_{-2} + q \alpha_{-1} + l \alpha_0 + m \alpha_1 + n \alpha_2 \in \Delta_{\mathbf{(A}_{2}\mathbf{)}_{-2}} \label{alphagen}, \qquad   p,q,l,m,n \in \mathbb{Z},
\end{equation}
to be 2, leads to the Diophantine equation
\begin{equation}
	l^2-l (m+n+q)+m^2-m n+n^2+p^2-p q+q^2 = 1 \qquad \Leftrightarrow \quad \alpha \cdot \alpha =2 . \label{Dioph}
\end{equation}
For every simple root $\alpha_i$ we define a Weyl reflection
 \begin{equation}
 	\sigma_i(x) := x- (\alpha_i \cdot x ) \alpha_i , \label{defWeyl}
 \end{equation}
about the hyperplane orthogonal to the root through the origin. Since Weyl reflections are orthogonal transformations, the inner product, $\alpha \cdot \alpha = \sigma_i(\alpha) \cdot \sigma_i(\alpha) $, is preserved. The symmetries of the Diophantine equation generated by the action of the Weyl reflections on $\alpha$ in this manner are
\begin{eqnarray}
&&\sigma_{-2}(\alpha) : p \rightarrow q-p , \quad 
	\sigma_{-1}(\alpha) : q \rightarrow l+p-q , \quad 
	\sigma_{0}(\alpha) : l \rightarrow q+m+n-l ,  \label{LWeylref} \\
	&&\sigma_{1}(\alpha) : m \rightarrow l+n-m , \quad 
	\sigma_{2}(\alpha) : n \rightarrow l+m-n . \notag
\end{eqnarray}
The composition of several different Weyl reflections can be used to define different types of Coxeter elements corresponding to the various subgroups and orderings of the Weyl reflections.

\subsection{Finite dimensional Coxeter orbits}

Coxeter elements $\sigma$ are in general defined as products of Weyl reflections involving all simple roots, i.e.  $\sigma:= \prod_{i=1}^r \sigma_i$ with $\alpha_i$ being simple and $r$ denoting the rank of the group. Since Weyl reflections do no commute  in general, different orderings in the product lead to different Coxeter elements, although they are still in the same equivalence class and have the same order. There are a number finite dimensional subgroups of the full Lorentzian Coxeter group that lead to versions of Coxeter elements of finite order that we will be briefly consider first for our example. 
\subsubsection{$A_2$ Coxeter orbits of order 3}
Commencing with the standard definition of an $A_2$-Coxeter element in the form
\begin{equation}
	\sigma_{A2}  :=  \sigma_1 \sigma_2 ,
\end{equation}
we easily calculate with (\ref{LWeylref}) its action on the generic $\mathbf{(A}_{2}\mathbf{)}_{-2}$ root $\alpha$  as defined in (\ref{alphagen})
\begin{eqnarray}
	\sigma_{A2}({\alpha}) &=& p \alpha _{-2} + q \alpha _{-1}  + l \alpha _0 + (2 l-n) \alpha _1  + (l+m-n) \alpha _2 , \notag \\ \notag
	\sigma_{A2}^2({\alpha}) &=& p \alpha _{-2} + q \alpha _{-1}  + l \alpha _0 + (l-m+n) \alpha _1 + (2 l-m) \alpha _2  , \\
		\sigma_{A2}^3({\alpha}) &=&  \alpha .
\end{eqnarray} 
Orbits with similar structure and in particular the same order $3$ arise when defining a Coxeter element by means of the product of Weyl reflections associated to any other two roots connected on the Dynkin diagram, e.g. $\tilde{\sigma}_{A2}  :=  \sigma_{-2} \sigma_{-1}$, $\hat{\sigma}_{A2}  :=  \sigma_{0} \sigma_{1}$ or $\bar{\sigma}_{A2}  :=  \sigma_{-1} \sigma_{0}$.

\subsubsection{$A_3$ Coxeter orbits of order 4}

A Coxeter element of order 4 can be constructed from three consecutive Weyl reflections associated to roots connected on the Dynkin diagram, thus corresponding to an $A_3$-diagram. For instance, defining 
\begin{equation}
	\sigma_{A3}  :=  \sigma_{-2} \sigma_0 \sigma_{-1} ,
\end{equation}
we compute
\begin{eqnarray}
	\sigma_{A3}({\alpha}) &=&(l-q) \alpha _{-2} + (l+p-q) \alpha _{-1} + (m+n+p-q)\alpha _0 + m \alpha _1 +n \alpha _2   , \notag \\  
	\sigma_{A3}^2({\alpha}) &=& (-l+m+n) \alpha _{-2} + (m+n-q) \alpha _{-1} + (m+n-p) \alpha _0  +m \alpha _1 +n \alpha _2   ,  \notag \\
	\sigma_{A3}^3({\alpha}) &=& (q-p) \alpha _{-2}  +(-l+m+n-p+q) \alpha _{-1} + (-l+m+n+q) \alpha _0 + m \alpha _1 +n \alpha _2   ,  \notag \\
	\sigma_{A3}^4({\alpha}) &=&  \alpha .
\end{eqnarray}  
Thus the Coxeter element $\sigma_{A3}$ has order 4.

\subsubsection{$A_4$ Coxeter orbits of order 5}

Similarly, when defining a Coxeter element from four  consecutive Weyl reflections associated to roots connected parts of the Dynkin diagram we obtain an $A_4$-diagram. In this manner we obtain one of finite order 5. For instance, defining 
\begin{equation}
	\sigma_{A4}  :=  \sigma_{-2} \sigma_0 \sigma_{-1} \sigma_{1} ,
\end{equation}
we compute
\begin{eqnarray}
	\sigma_{A4}({\alpha}) &=& (l - q) \alpha _ {-2} + (l + p - q)  \alpha _ {-1}  + (l - m + 2 n + 
	p - q) \alpha _ 0 + (l - m + n) \alpha _ 1 + n \alpha _ 2   , \notag \\ \notag
	\sigma_{A4}^2({\alpha}) &=&  (2 n - m) \alpha _ {-2} + (l - m + 2 n - q) \alpha _ {-1} + (3 n - 
	q) \alpha _ 0  + (2 n + p - q) \alpha _ 1 + n \alpha _ 2   , \\\notag
	\sigma_{A4}^3({\alpha}) &=&  (m + n - l) \alpha _ {-2} + (3 n - l)\alpha _ {-1} + (3 n - p + q - 
	l) \alpha _ 0 + (2 n - p) \alpha _ 1  + n \alpha _ 2   , \\\notag
	\sigma_{A4}^4({\alpha}) &=&  (q - p) \alpha _ {-2}  + (m + n - p + q - l) \alpha _ {-1} + (m + n + 
	q - l) \alpha _ 0 + (2 n + q - l )\alpha _ 1 + n \alpha _ 2   , \\  
	\sigma_{A4}^5({\alpha}) &=&  \alpha .
\end{eqnarray} 
Hence $\sigma_{A4}$ has order 5.

\subsubsection{$A_1 \otimes A_2$ Coxeter orbits of order 6}

Naturally we may also define Coxeter elements associated to roots corresponding to disconnected Dynkin diagrams, which lead to direct product of the subgroups.  For instance, defining 
\begin{equation}
	\sigma_{A12}  :=  \sigma_{1} \sigma_{-2} \sigma_{-1}  ,
\end{equation}
leads to orbits of order 6, associated to  $A_1 \otimes A_2$. We compute
\begin{eqnarray}
	\sigma_{A12}({\alpha}) &=&   (l - q) \alpha _ {-2} + (l + p - q) \alpha _ {-1} + 
	l \alpha _ 0 + (l - m + n) \alpha _ 1 + n \alpha _ 2  , \notag \\ \notag
	\sigma_{A12}^2({\alpha}) &=& (q - p) \alpha _ {-2}  + (l - p) \alpha _ {-1}  + l \alpha _ 0 + 
	m \alpha _ 1 + n \alpha _ 2   , \\\notag
	\sigma_{A12}^3({\alpha}) &=&  p \alpha _ {-2}  + q \alpha _ {-1}  + 
	l \alpha _ 0 + (l - m + n) \alpha _ 1 + n \alpha _ 2   , \\\notag
	\sigma_{A12}^4({\alpha}) &=&  ( l - q) \alpha _ {-2} + (l + p - q) \alpha _ {-1} + l\alpha _ 0 + 
	m \alpha _ 1 + n\alpha _ 2   , \\  \notag
	\sigma_{A12}^5({\alpha}) &=&  (q - p)\alpha _ {-2} + (l - p) \alpha _ {-1} + 
	l \alpha _ 0 + (l - m + n)\alpha _ 1 + n \alpha _ 2    , \\  
	\sigma_{A12}^6({\alpha}) &=&  \alpha .
\end{eqnarray} 
The order $6$ of this Coxeter element simply results from multiplying the two Coxeter numbers of $A_1$ and $A_2$, i.e. $h_{A_1} \times h_{A_2} = 2 \times 3$.

\subsection{Infinite dimensional $\mathbf{(A}_{2}\mathbf{)}_{0}$-affine Kac-Moody Coxeter orbits}
We now come to the cases of our main interest, which is the construction of infinite Weyl groups that consequently possess Coxeter elements of infinite order. The simplest example for such a group is the affine extension of the $A_2$ Lie algebra, see e.g. \cite{kacinfinite}. We will show in detail how the generic formula for the action of an arbitrary power of the Coxeter element on a root can be derived. Defining a specific Coxeter element for the $\mathbf{(A}_{2}\mathbf{)}_{0}$-affine Kac-Moody algebra as
\begin{equation}
	\sigma_a  :=  \sigma_0 \sigma_1 \sigma_2 , \label{affCox}
\end{equation}
we calculate its orbits. When acting consecutively with $\sigma_a$ on the arbitrary root $\alpha$ in (\ref{alphagen}) we compute
\begin{eqnarray}
	\sigma_a({\alpha}) &=& p \alpha _{-2} + q \alpha _{-1} + (2 l+m-2 n+q) \alpha _0 +(2 l-n) \alpha _1 + (l+m-n) \alpha _2 ,  \\ 
	\sigma_a^2({\alpha}) &=& p \alpha _ {-2} + 
	q \alpha _ {-1} + (4 l - 3 n + 3 q)\alpha _ 0 + (3 l + m - 3 n + 
	2 q)\alpha _ 1 + (3 l - 2 n + q)\alpha _ 2 .\notag
\end{eqnarray} 
For arbitrary powers we obtain 
\begin{equation}
	\sigma_a^{ k}(\alpha)  := p \alpha _ {-2} + 
	q \alpha _ {-1} + \sum_{\nu=0}^{2} \sum_{\mu=p,q,l,m,n}  a_\nu^\mu(k) \mu \alpha_\nu,   \label{siga02k}
\end{equation} 
where the challenge is to determine the coefficients $a_\nu^\mu(k)$ for any power  $k \in \mathbb{Z}$. When acting with $\sigma$ on the generic expressions in (\ref{siga02k}) we 
easily identify the recurrence relations that need to be satisfied 
\begin{eqnarray}
	a_{0}^\mu(k+1)	&=&  a_{-1}^\mu(k) + 2 a_0^\mu(k) + a_1^\mu(k) - 2 a_2^\mu(k), \label{reca01} \\
	a_{1}^\mu(k+1)	&=& 2   a_0^\mu(k) - a_2^\mu(k) ,\\
	a_{2}^\mu(k+1)	&=&  a_0^\mu(k) + a_1^\mu(k) - a_2^\mu(k). \label{reca05} 
\end{eqnarray}
Thus we may write these equations more compactly as 
\begin{equation}
a_{\nu}^\mu(k+1) = \left( M  a^\mu(k) \right)_\nu  = \left( M^{k+1}  a^\mu(0) \right)_\nu ,
\end{equation}
involving the matrix and the vector 
\begin{equation}
  M= \left(
	\begin{array}{ccccc}
		1 & 0 & 0 & 0 & 0 \\
		0 & 1 & 0 & 0 & 0 \\
		0 & 1 & 2 & 1 & -2 \\
		0 & 0 & 2 & 0 & -1 \\
		0 & 0 & 1 & 1 & -1 \\
	\end{array}
	\right), \quad \text{and} \quad a^\mu(k) = [a_{-2}^\mu(k),a_{-1}^\mu(k),a_{0}^\mu(k),a_{1}^\mu(k),a_{2}^\mu(k) ] , \label{matrepr}
\end{equation}
respectively. Taking as initial condition the root $\alpha$ as defined in (\ref{alphagen}), we have
\begin{equation}
	a_{-2}^p(0)=a_{-1}^q(0)=a_{0}^l(0)=a_{1}^m(0)=a_{2}^n(0)=1, \quad \text{and} \quad  a_\nu^\mu(0)=0 \,\, \text{otherwise}, \label{ini111}
\end{equation} 
so that the general solution is 
\begin{equation}
     a_{\nu}^\mu(k)= ( M^k)_{\nu \mu}  . \label{matsol}
\end{equation}
In order to compute the powers of the matrix $M$ we may assume that the coefficient functions satisfy a recurrence relation of the general form
\begin{equation}
	a_{\nu}^\mu(k+1) = \sum_{i=1}^N c_i a_{\nu}^\mu(k+1-i) ,  \label{matrep}
\end{equation}
with some constants $c_i \in \mathbb{C}$ and $N \in \mathbb{N}$ that need to be determined. See for instance \cite{epp2010discrete} for a treatment of such type of recurrence relations and for general techniques to solve them. Employing the matrix representation (\ref{matrepr}) we find for $N=4$ indeed the following recurrence relation to hold
\begin{equation}
	a_{\nu}^\mu(k+1) = 2 a_{\nu}^\mu(k) - 2 a_{\nu}^\mu(k-2) + b_{\nu}^\mu(k-3) .  \label{recurkm11}
\end{equation}
The corresponding fourth order characteristic equation for this relation resulting from $ 	a_{\nu}^\mu(k) \sim x^k $ is 
\begin{equation}
	x^4 - 2 x^3  + 2 x -1 =0.
\end{equation}
The four roots of this equation are found to be
\begin{equation}
	\lambda_1 =-1, \quad \lambda_{2/3/4} =1 .
\end{equation}
As we have a threefold degenerate solution to the chracteristic equation, the general solution to the recurrence relation (\ref{recurkm11}) is 
\begin{equation}
	a_{\nu}^\mu(k) = \tilde{c}_1 (-1)^k +  \tilde{c}_2 +  \tilde{c}_3 k +  \tilde{c}_4  k^2,   \label{rec3411}
\end{equation}
where the constant coefficients $\tilde{c}_i$, $i=1,\ldots, 4$ are determined from the explicit evaluation of the lowest powers of the matrix in (\ref{matsol}). In this manner we find the generic formula
\begin{eqnarray}
\sigma_a^{k}(\alpha)&=& p\alpha _{-2} +q \alpha _{-1} \label{Coxeteraffgen} \\
&& \!\!\! \!\!\!\!\!\! \!\!\!+ \left[\frac{1}{8} \left(6 k^2+1\right) q+\frac{1}{4} (6 k+3) l-\frac{1}{4} (6 k+1)
n+\frac{m}{2}+(-1)^k\frac{2l-4m+2n-q}{8} \right] \alpha _0  \notag  \\ 
&& \!\!\! \!\!\!\!\!\! \!\!\! +  
\left[\frac{1}{8} \left(6 k^2-4 k-1\right) q+\frac{1}{4} (6 k+1) l-\frac{1}{4} (6 k-1)
n+\frac{m}{2}-(-1)^k \frac{2l-4m+2n-q}{8} \right]\alpha _1  \notag\\
&& \!\!\! \!\!\! \!\!\! \!\!\!+ \left[   \frac{1}{8} \left(6 k^2-8 k+1\right) q+\frac{1}{4} (6 k-1) l-\frac{1}{4} (6 k-3)
n+\frac{m}{2} +(-1)^k \frac{2l-4m+2n-q}{8}  \right] \alpha _2.  
   \notag
\end{eqnarray} 
Thus, as expected, $\sigma_a$ defined in equation (\ref{affCox}) is a Coxeter element of infinite order.

\subsection{Infinite dimensional $\mathbf{(A}_{2}\mathbf{)}_{-1}$-hyperbolic Kac-Moody Coxeter orbits}
Besides affine extensions of finite dimensional algebras \cite{kacinfinite} also the hyperbolic algebras have been fully classified \cite{carbone2010}. These algebras can be defined based on their connected Dynkin diagrams, which exhibit the property that the removal of {\emph{any}} node results in a set of connected Dynkin diagrams, possibly disconnected, all of which are of finite type except for at most one which may be of affine type. This is easily checked to be true for the Dynkin diagram at the beginning of section 2 with $\alpha_{-2}$ removed. 

The simplest example for such an algebra is the extended affine $A_2$-algebra, i.e. the $\mathbf{(A}_{2}\mathbf{)}_{-1}$-hyperbolic Kac-Moody algebra for which we define a specific Coxeter element as
\begin{equation}
	\sigma_h  := \sigma_{-1} \sigma_0 \sigma_1 \sigma_2 , \label{hypCox}
\end{equation}
and calculate its infinite dimensional orbits. Acting on the arbitrary $\mathbf{(A}_{2}\mathbf{)}_{-2}$ root $\alpha$ in (\ref{alphagen}) we compute the first consecutive actions as 
\begin{eqnarray}
	\sigma_h({\alpha}) &=& p\alpha _{-2} + (2 l+m-2 n+p) \alpha _{-1}+ (2 l+m-2 n+q) \alpha _0 + (2 l-n) \alpha _1   \qquad \qquad \\ 
	&&  + (l+m-n) \alpha _2, \notag\\
	\sigma_h^{2}(\alpha) & =& p \alpha _{-2} + (4 l-3 n+p+2 q) \alpha _{-1}+   (6 l+m-5 n+p+2 q) \alpha _0  \\
	&& + (3 l+m-3 n+2 q) \alpha _1 + (3 l-2 n+q) \alpha _2  ,\notag
\end{eqnarray} 
etc. For arbitrary powers we assume a similar expansion as in (\ref{siga02k}) of the previous section 
\begin{equation}
	\sigma_h^{k}(\alpha)  := p \alpha _{-2} + \sum_{\nu=-1}^{2} \sum_{\mu=p,q,l,m,n} b_\nu^\mu(k) \mu \alpha_\nu,   \label{sigkm2k}
\end{equation} 
where $b_\nu^\mu(k),k \in \mathbb{Z}$. Acting with $\sigma_h$ on the generic expansion in (\ref{sigkm2k}) we identify the recurrence relations to be satisfied by the coefficients $ b_\nu^\mu(k)$
\begin{eqnarray}
	b_{-1}^\mu(k+1)	&=& b_{-2}^\mu(k) + 2 b_0^\mu(k) +  b_1^\mu(k)-2 b_2^\mu(k), \\
	b_{0}^\mu(k+1)	&=&  b_{-1}^\mu(k) + 2 b_{0}^\mu(k) + b_1^\mu(k) -  2 b_2^\mu(k),  \\
	b_{1}^\mu(k+1)	&=& 2  b_{0}^\mu(k) -  b_2^\mu(k) ,\\
	b_{2}^\mu(k+1)	&=& b_{0}^\mu(k) +b_1^\mu(k) -  b_2^\mu(k). \label{kmrec5}
\end{eqnarray}
Again we may cast the recurrence relations into matrix form as 
\begin{equation}
	b_{\nu}^\mu(k+1) = \left( \hat{M}  b^\mu(k) \right)_\nu  = \left( \hat{M}^{k+1}  b^\mu(0) \right)_\nu 
\end{equation}
involving the matrix and the vector 
\begin{equation}
	\hat{M}= \left(
	\begin{array}{ccccc}
		1 & 0 & 0 & 0 & 0 \\
		1 & 0 & 2 & 1 & -2 \\
		0 & 1 & 2 & 1 & -2 \\
		0 & 0 & 2 & 0 & -1 \\
		0 & 0 & 1 & 1 & -1 \\
	\end{array}
	\right), \quad \text{and} \quad b^\mu(k) = [b_{-2}^\mu(k),b_{-1}^\mu(k),b_{0}^\mu(k),b_{1}^\mu(k),b_{2}^\mu(k) ] , \label{matreprb}
\end{equation}
respectively. Taking $\alpha$ to be of the form (\ref{alphagen}) the initial conditions are
\begin{equation}
	b_{-2}^p(0)=bc_{-1}^q(0)=b_{0}^l(0)=b_{1}^m(0)=b_{2}^n(0)= 1, \quad \text{and} \quad  b_\nu^\mu(0)=0 \,\, \text{otherwise}.
\end{equation} 
As in the previous section we may consecutively calculate all the coefficients $ b_{\nu}^\mu(k)$ with some given initial condition. We also assume here that these coefficients satisfy a linear recurrence relation of the form
\begin{equation}
	b_{\nu}^\mu(k+1) = \sum_{i=1}^N c_i b_{\nu}^\mu(k+1-i) ,
\end{equation}
with constants $c_i \in \mathbb{C}$ and $N \in \mathbb{N}$ that need to be determined. For $N=5$ we find that indeed the following recurrence relation holds for all coefficients
\begin{equation}
	b_{\nu}^\mu(k+1) = 2 b_{\nu}^\mu(k) + 2 b_{\nu}^\mu(k-1)- 2 b_{\nu}^\mu(k-2) -2 b_{\nu}^\mu(k-3) + b_{\nu}^\mu(k-4).  \label{recurkm}
\end{equation}
Following the same procedure as in the previous section we determine the fifth order characteristic polynomial for (\ref{recurkm}), by assuming $ 	b_{\nu}^\mu(k) \sim x^k $. The characteristic equation then reads
\begin{equation}
	x^5 - 2 x^4  - 2 x^3 +2  x^2 + 2 x -1 =0. \label{charpol}
\end{equation}
We find three real and two complex roots to equation (\ref{charpol}) 
\begin{equation}
	\lambda_1 =1, \quad \lambda_{2/3} =\frac{1}{4} \left[1+\sqrt{21}\pm\sqrt{ 2\sqrt{21} + 6}\right] , \quad   \lambda_{4/5} =\frac{1}{4} \left[1-\sqrt{21} \pm i \sqrt{ 2\sqrt{21}-6}\right] .
\end{equation}
Thus according to the general theory of recurrence relations, see e.g. \cite{epp2010discrete}, the solution to equation (\ref{recurkm}) can be cast into the form 
\begin{equation}
	b_{\nu}^\mu(k) =  b_{\mu 1}^\nu \lambda_1^k +  b_{\mu 2}^\nu \lambda_2^k +  b_{\mu 3}^\nu \lambda_3^k +  b_{\mu 4}^\nu \lambda_4^k + b_{\mu 5}^\nu \lambda_5^k,   \label{rec34}
\end{equation}
where the constant coefficients $b_{\mu i}^\nu$, $i=1,\ldots, 5$ are determined from the initial conditions $b_{\nu}^\mu(0)$ and the first four iterations that give $b_{\nu}^\mu(1) \ldots, b_{\nu}^\mu(4)$ for different $\nu$ and $\mu$. To stress the reality of these solutions we can equivalently express the coefficients as 
\begin{equation}
	b_{\nu}^\mu(k) =  b_{\mu 1}^\nu +  \hat{b}_{\mu 2}^\nu \Lambda_2^k +  \hat{b}_{\mu 3}^\nu \Lambda_3^k +  \hat{b}_{\mu 4}^\nu \Lambda_4^k + \hat{b}_{\mu 5}^\nu \Lambda_5^k,  ,   \label{rec34}
\end{equation}
where $b_{\mu 2/3}^\nu=\hat{b}_{\mu 3}^\nu \pm \hat{b}_{\mu 2}^\nu$,  
$b_{\mu 4/5}^\nu=  (\hat{b}_{\mu 5}^\nu \mp \hat{b}_{\mu 4}^\nu)/2$ and
\begin{equation}
	\Lambda_{2/3}^k=\frac{1}{\sqrt{21}}(\lambda_2^k \mp \lambda_3^k), \qquad 
	\Lambda_4^k=\frac{1}{2\sqrt{21}} i(\lambda_5^k-\lambda_4^k) \qquad
	\Lambda_5^k=\frac{1}{2\sqrt{21}}(\lambda_5^k+\lambda_4^k). 
\end{equation}
Using the initial values as stated above we find the solutions
\begin{eqnarray}
	b^p_{-1}&=&\Lambda _3^k-2 \Lambda _5^k \\
	b^q_{-1}&=&\frac{1}{2}[\mu _- (\mu _- \Lambda _3^k+2 \Lambda
	_4^k)+\mu _+ (\Lambda _2^k+2 \mu _+ \Lambda
	_5^k)] \\
	b^l_{-1}&=&\frac{1}{2}[\varphi _+ \Lambda _2^k+3 \Lambda _3^k-2 \varphi
	_- \Lambda _4^k-6 \Lambda _5^k]\\  
	b^m_{-1}&=&\frac{1}{6}[\sqrt{3}  \mu _-^3\Lambda _2^k+3 \Lambda _3^k+2
	\sqrt{3} \mu _+^3 \Lambda _4^k-6 \Lambda _5^k]\\
	b^n_{-1}&=&-\mu _+ \Lambda _2^k-\Lambda _3^k-2 \mu _- \Lambda _4^k+2 \Lambda
	_5^k  \\
	&&  \notag \\
	b^p_{0}&=&-1-\frac{1}{2}
	\mu _+ \Lambda _2^k+\frac{1}{6} \mu _+^4 \Lambda _3^k-\mu _-\Lambda _4^k  -\frac{1}{3}\mu _-^4 \Lambda _5^k \\
	b^q_{0}&=& \mu _+ \Lambda _2^k+2 \mu _- \Lambda _4^k \\
	b^l_{0}&= &\frac{1}{6}[\mu _+^2 (\sqrt{3} \varphi _- \Lambda _2^k+3
	\Lambda _3^k)-2 \mu _-^2 (\sqrt{3} \varphi _+ \Lambda
	_4^k-3 \Lambda _5^k)]  \qquad \qquad \\  
	b^m_{0}&=& \Lambda _3^k-2 \Lambda _5^k \\
	b^n_{0}&=&-\frac{1}{2} \mu _+^3 \Lambda
	_2^k-\frac{1}{2}\Lambda _3^k+ \mu _-^3 \Lambda _4^k +\Lambda _5^k
\end{eqnarray}

\begin{eqnarray}
	b^p_{1}&=&-1+\frac{1}{6}[-\sqrt{21} \mu _+ \Lambda _2^k+\mu _+^4 \Lambda
	_3^k+2 \sqrt{21} 
	\mu _-\Lambda _4^k-2  \mu _-^4\Lambda _5^k] \qquad \qquad \\
	b^q_{1}&=&\frac{1}{2} \varphi _+ \Lambda _2^k-\frac{1}{2}\Lambda _3^k-\varphi
	_- \Lambda _4^k+\Lambda _5^k \\
	b^l_{1}&=&  \mu _+ \Lambda _2^k+\Lambda _3^k+2 \mu _- \Lambda _4^k-2 \Lambda
	_5^k \\  
	b^m_{1}&=& -\frac{1}{2 \sqrt{3}}(\mu _- \Lambda
	_2^k-2\mu _+ \Lambda
	_4^k)+ \frac{ 1}{2 \sqrt{21}}(\theta _-^2 \Lambda _3^k+2\theta _+^2\Lambda _5^k) \\
	b^n_{1}&=&-\frac{1}{6}
	\tau _- \mu _+^3 \Lambda _2^k+\frac{1}{2}\Lambda _3^k -\frac{\tau _+  \mu _-^3}{3 \sqrt{21}}\Lambda _4^k-\Lambda
	_5^k \\
	&&  \notag  \\
	b^p_{2}&=&-1+\frac{1}{6}[-\mu _+^2 (\varphi _+
	\Lambda _2^k-\sqrt{21} \Lambda _3^k)-2 \mu _-^2 (\varphi _- \Lambda
	_4^k+\sqrt{21} \Lambda _5^k)] \\
	b^q_{2}&= &\mu _+ \Lambda _2^k-\Lambda _3^k+2 \Lambda _5^k+\frac{2 \mu _-}{\sqrt{21}} \Lambda _4^k \\
	b^l_{2}&= &\frac{1}{2} \varphi _+ \Lambda _2^k+\frac{1}{2}\Lambda _3^k-\varphi
	_- \Lambda _4^k-\Lambda _5^k  \\  
	b^m_{2}&=&-\frac{\mu _-}{3 \sqrt{7}}\Lambda _2^k +\Lambda _3^k+\frac{2 \mu _+}{\sqrt{3}}\Lambda _4^k-2 \Lambda
	_5^k \\
	b^n_{2}&=&-\frac{1}{126}[\theta _+^2 (\mu _+^3
	\Lambda _2^k-3 \Lambda _3^k)+2 \theta _-^2 (\mu _-^3 \Lambda _4^k-3
	\Lambda _5^k)]
\end{eqnarray}
We abbreviated here some of the lengthy square root constants to achieve a more compact presentation
\begin{eqnarray}
	\mu_\pm&:=&\sqrt{\frac{1}{2}\left(\sqrt{21}\pm3\right)}, \quad \mu_\pm:=\sqrt{2\sqrt{21}\pm3}, \\
   \theta_\pm&:=&\sqrt{\frac{1}{2}(21\pm\sqrt{21})}, \quad \tau_\pm:=\sqrt{25\pm4\sqrt{21}} .
\end{eqnarray}
Evidently the Coxeter element $\sigma_h$ as defined in (\ref{hypCox}) is of infinite order.

\subsection{Infinite dimensional $\mathbf{(A}_{2}\mathbf{)}_{-2}$-Lorentzian Kac-Moody Coxeter orbits}
Adding a further node to the $(A_2)_{-1}$-diagram at the $\alpha_{-1}$-node leads to a diagram that no longer satisfies the classification criteria for a hyperbolic Kac-Moody algebra. This is because when removing the $\alpha_{-2}$-node one is not left with a set of disconnected diagrams of finite type except for at most one affine type. The diagram obtained by removing this node is still of hyperbolic type. Thus the new diagram as depicted at the beginning of section 2 is of a new type, referred to as Lorentzian. To characterise these diagrams it was proposed in \cite{gaberdiel2002class} to relax the defining criterion for the hyperbolic Kac-Moody algebra in the sense that the required decomposition must not be obtained for all nodes but only {\emph{for at least one}}. Our $(A_2)_{-2}$-example clearly satisfies this requirement as it even holds for all nodes except for the $\alpha_{-2}$-node. 

We define a Coxeter element for this algebra in the usual manner by the product of Weyl reflections associated to each simple root in a particular order
 \begin{equation}
	\sigma_L  := \sigma_{-2} \sigma_{-1} \sigma_0 \sigma_1 \sigma_2 . \label{CoxLor}
\end{equation}
We calculate its orbits by acting with $\sigma_L$ consecutively on the root $\alpha$ in (\ref{alphagen}), computing 
\begin{eqnarray}
	\sigma_L({\alpha}) &=&  (2 l + m - 2 n)\alpha _ {-2} + (2 l + m - 2 n + 
	p)\alpha _ {-1} + (2 l + m - 2 n + q)\alpha _ 0  \\ 
	   &&+ (2 l - 
	n)\alpha _ 1 + (l + m - n)\alpha _ 2  \notag\\
	\sigma_L^{2}(\alpha) & =&  (4 l - 3 n + 2 q)\alpha _ {-2} + (6 l + m - 5 n + 
	2 q) \alpha _ {-1} + (6 l + m - 5 n + p + 
	2 q) \alpha _ 0  , \quad \\
	&& + (3 l + m - 3 n + 2 q) \alpha _ 1 + (3 l - 2 n +
	q) \alpha _ 2 .\notag
\end{eqnarray} 
For arbitrary powers we make a similar assumption as in the previous sections 
\begin{equation}
	\sigma_L^{k}(\alpha)  := \sum_{\nu=-2}^{2} \sum_{\mu=p,q,l,m,n} c_\nu^\mu(k) \mu \alpha_\nu,   \label{sig2k}
\end{equation} 
with unknown constants $c_\nu^\mu(k),k \in \mathbb{Z}$. Acting with $\sigma_L$ on the generic expressions in (\ref{sig2k}) we 
easily identify the recurrence relations to be satisfied by the coefficients $ c_\nu^\mu(k)$
\begin{eqnarray}
   c_{-2}^\mu(k+1)	&=&  2 c_{0}^\mu(k) + c_1^\mu(k) - 2 c_2^\mu(k) , \label{rec1} \\
   c_{-1}^\mu(k+1)	&=& c_{-2}^\mu(k) + 2 c_0^\mu(k) +  c_1^\mu(k)-2 c_2^\mu(k), \\
  c_{0}^\mu(k+1)	&=&  c_{-1}^\mu(k) + 2 c_{0}^\mu(k) + c_1^\mu(k) -  2 c_2^\mu(k),  \\
   c_{1}^\mu(k+1)	&=& 2  c_{0}^\mu(k) -  c_2^\mu(k) ,\\
    c_{2}^\mu(k+1)	&=& c_{0}^\mu(k) +c_1^\mu(k) -  c_2^\mu(k). \label{rec5}
\end{eqnarray}
Also in this case a more compact form of the recurrence relations can be achieved in a matrix formulation
\begin{equation}
	c_{\nu}^\mu(k+1) = \left( \tilde{M}  c^\mu(k) \right)_\nu  = \left( \tilde{M}^{k+1}  c^\mu(0) \right)_\nu 
\end{equation}
involving the matrix and the vector 
\begin{equation}
	\tilde{M}= \left(
	\begin{array}{ccccc}
		0 & 0 & 2 & 1 & -2 \\
		1 & 0 & 2 & 1 & -2 \\
		0 & 1 & 2 & 1 & -2 \\
		0 & 0 & 2 & 0 & -1 \\
		0 & 0 & 1 & 1 & -1 \\
	\end{array}
	\right), \quad \text{and} \quad c^\mu(k) = [c_{-2}^\mu(k),c_{-1}^\mu(k),c_{0}^\mu(k),c_{1}^\mu(k),c_{2}^\mu(k) ] , \label{matreprbLor}
\end{equation}
respectively. With $\alpha$ in (\ref{alphagen}) the initial conditions are
\begin{equation}
c_{-2}^p(0)=c_{-1}^q(0)=c_{0}^l(0)=c_{1}^m(0)=c_{2}^n(0)= 1, \quad \text{and} \quad  c_\nu^\mu(0)=0 \,\, \text{otherwise}.
\end{equation} 
We assume once more that these coefficients satisfy a linear recurrence relation of the form
\begin{equation}
	c_{\nu}^\mu(k+1) = \sum_{i=1}^N \tilde{c}_i c_{\nu}^\mu(k+1-i) ,
\end{equation}
with some constants $\tilde{c}_i \in \mathbb{R}$ and $N \in \mathbb{N}$ that need to be determined. For $N=5$ we find that the following recurrence relation holds for all coefficients
\begin{equation}
	c_{\nu}^\mu(k+1) =  c_{\nu}^\mu(k) + 3 c_{\nu}^\mu(k-1)+ 3 c_{\nu}^\mu(k-2) + c_{\nu}^\mu(k-3) - c_{\nu}^\mu(k-4).  \label{recurLL}
\end{equation}
The fifth order characteristic equation for this relation resulting from $ 	c_{\nu}^\mu(k) \sim x^k $ is 
\begin{equation}
	x^5 - x^4  - 3 x^3 - 3 x^2 - x +1 =0.
\end{equation}
Similarly as for the hyperbolic case we find three real and two complex roots to this equation
\begin{equation}
\lambda_1 =-1,  \quad   \lambda_{2/3} =\frac{1}{2} \left(  3 \pm  \sqrt{5}   \right) = \phi^{\pm 2}, \quad \lambda_{4/5} =\omega^{\pm 1},
\end{equation}
where $\omega$ is a third root of unity $\omega:= \exp( 2 \pi i /3)$  and $\phi$ is the golden ratio $\phi:= (1 + \sqrt{5})/2$ with negative inverse golden ratio $\bar{\phi}:= (1 - \sqrt{5})/2$. Thus the general solution to the recurrence relation (\ref{recurLL}) is 
\begin{equation}
	c_{\nu}^\mu(k) = c_1 \lambda_1^k +  c_2 \lambda_2^k +  c_3 \lambda_3^k +  c_4 \lambda_4^k + c_5 \lambda_5^k,   \label{rec34}
\end{equation}
where the constant coefficients $c_i$, $i=1,\ldots, 5$ are determined from the initial conditions $c_{\nu}^\mu(0)$ and the first four explicit iterations $ c_{\nu}^\mu(1) \ldots, c_{\nu}^\mu(4)$ for different $\nu$ and $\mu$.  Performing this calculation we find the solutions
\begin{eqnarray}
  	  	 	\tilde{c}_{-2}^m (k)    &=&  L_{2 k+2}+10 \sin \left(\frac{\pi }{6}+\frac{2 \pi  k}{3}\right)-8 (-1)^k  \\
  	  	\tilde{c}_{-2}^n (k) &=& L_{2 k}-L_{2 k+5}+10 \sin \left(\frac{2 \pi  k}{3}-\frac{\pi }{6}\right)+4(-1)^k,  \\
  	  	 \tilde{c}_{-2}^p (k) &=&  L_{2 k+1}+10 \sqrt{3} \cos \left(\frac{2 \pi  k}{3}-\frac{\pi }{6}\right)+4 (-1)^k,  \\
  	  	  &&  \notag \\	  		
  	     	\tilde{c}_{-1}^m (k) &=& 2 L_{2 k}+L_{2 k+1}-10 \sin \left(\frac{2 \pi  k}{3}-\frac{\pi }{6}\right),  \\
  	    \tilde{c}_{-1}^n (k) &=& -L_{2 k}-8 L_{2 k+1}+10 \cos \left(\frac{2 \pi  k}{3}\right)  ,   \\
  	      \tilde{c}_{-1}^p (k) &=& -L_{2 k}+2 L_{2 k+1}+10 \sqrt{3} \sin \left(\frac{2 \pi  k}{3}\right)  ,  \\
  	     &&  \notag \\	 	    
  	        	\tilde{c}_{0}^m (k) &=&4 \left(L_{2 k}-2\right) = F_k^2,  \\
  	    \tilde{c}_{0}^n (k) &=& 4 \left(L_{2 k}-3 L_{2 k+1}+1\right) =F_k  F_{k-1} ,   \\
  	       \tilde{c}_{0}^p (k) &=& 4 \left(L_{2 k-1}+1\right) =F_k (F_{k}-3 F_{k+1})  ,  \\
  	      &&  \notag \\	
  	      \tilde{c}_{1}^m (k) &=&4 L_{2 k}-L_{2 k+1}+\frac{10}{\sqrt{3}} \cos \left(\frac{2 \pi  k}{3}+\frac{\pi }{6}\right)+8(-1)^k ,  \\
  	      \tilde{c}_{1}^n (k) &=& 7 L_{2 k}-10 L_{2 k+1}-\frac{10}{\sqrt{3}} \sin \left(\frac{2 \pi  k}{3}\right)-4(-1)^k  ,  \\
  	      \tilde{c}_{1}^p (k) &=& 4 L_{2 k+1}-5 L_{2 k}+10 \cos \left(\frac{2 \pi  k}{3}\right)-4(-1)^k   ,  
  	  \end{eqnarray}
 \begin{eqnarray}	
 	\tilde{c}_{2}^m (k) &=&5 L_{2 k}-2 L_{2 k+1}+\frac{10}{\sqrt{3}} \sin \left(\frac{2 \pi  k}{3}\right)-8 (-1)^k,  \\
 	\tilde{c}_{2}^n (k) &=& -11 L_{2 k-1}+\frac{10}{\sqrt{3}} \cos \left(\frac{\pi }{6}-\frac{2 \pi  k}{3}\right)+4 (-1)^k ,  \\
 	\tilde{c}_{2}^p (k) &=& 5 L_{2 k+1}-7 L_{2 k}+10 \sin \left(\frac{\pi }{6}+\frac{2 \pi  k}{3}\right)+4(-1)^k   ,     \\     
  	     && \notag \\
  	      \tilde{c}_{\nu}^p (k) &=&\tilde{c}_{\nu}^q (k-1) =\tilde{c}_{\nu}^l (k-2),
\end{eqnarray}
where $c_\nu^\mu(k) = 1/20 \tilde{c}_\nu^\mu(k) $, $F_n$ label the Fibonacci numbers with $n \in \mathbb{N}$ and $L_n$ are Lukas numbers  $L_n:= \phi^n + \bar{\phi}^{n}$. The Coxeter element $\sigma_L$ as defined in (\ref{CoxLor}) is therefore of infinite order.

Our next aim here is to utilise the mathematical structures discussed in this section in the context of some physical models. In particular we will formulate Calogero type systems that respect these infinite symmetries. An important question to address is whether the models based on infinite dimensional root spaces are integrable? As in the finite case, the existence and knowledge of the polynomial invariants largely facilitates to answer that question. One may attempt to construct them from scratch, but there is an elegant way to obtain them for Weyl groups associated to Dynkin diagrams that can be bicoloured \cite{carter1989simple,dorey1991root,Mass2}, that is to associate one of two colours to each node in the Dynkin diagram with no two adjacent nodes of the same colour. Unfortunately a bicolouration is not possible for our simple example of the $(A_2)_{-2}$-diagram. Therefore let us briefly discuss the structure of the invariants for the $(A_3)_{-2}$-Lorentzian Kac-Moody algebra.

\section{$\mathbf{(A}_{3}\mathbf{)}_{-2}$-Lorentzian Kac-Moody algebra}
The Cartan matrix $K$ for the $(A_3)_{-2}$-Lorentzian Kac-Moody algebra and the corresponding Dynkin diagram are\\
\setlength{\unitlength}{0.58cm} 
\begin{picture}(13.00,5.)(0.0,2.5)
	\thicklines
	\put(1.5,5.0){\Large{ $\mathbf{(A}_{3}\mathbf{)}_{-2}:$ }}
	\put(5.3,6){{$\small{\alpha_{1}}$}}
	\put(5.3,5){{$\small{\alpha_{2}}$}}
	\put(5.3,4){{$\small{\alpha_{3}}$}}
	\put(6.2,4.2){\line(0,1){0.9}}
	\put(6.2,5.35){\line(0,1){0.9}}
	\put(6.0,5){\Large{$\circ$}}
	\put(6.0,6){\Large{$\bullet$}}
	\put(6.0,4){\Large{$\bullet$}}
	\put(7.1,5.3){\line(-1,1){0.8}}
	\put(7.1,5.1){\line(-1,-1){0.8}}
	\put(7.0,5){\Large{$\circ$}}
	\put(7.3,5.2){\line(1,0){0.8}}
	\put(6.8,4.5){{$\small{\alpha_{0}}$}}
	\put(8.0,5){\Large{$\bullet$}}
	\put(8.35,5.2){\line(1,0){0.78}}
	\put(7.8,4.5){{$\small{\alpha_{-1}}$}}
	\put(9.0,5){\Large{$\circ$}}
	\put(9.0,4.5){{$\small{\alpha_{-2}}$}}
	\put(13.0,5){$ \!\!\!  \!\!\! \!\!\! K_{ij}=2 \frac{\alpha_i \cdot \alpha_j}{\alpha_j \cdot \alpha_j}=\left(
		\begin{array}{cccccc}
			2 & -1 & 0 & 0 & 0 & 0 \\
			-1 & 2 & -1 & 0 & 0 & 0 \\
			0 & -1 & 2 & -1 & 0 & -1 \\
			0 & 0 & -1 & 2 & -1 & 0 \\
			0 & 0 & 0 & -1 & 2 & -1 \\
			0 & 0 & -1 & 0 & -1 & 2 \\
		\end{array}
		\right)_{ij}.$}
\end{picture}\\
Clearly this diagram is of Lorentzian type as for instance the removal of the $\alpha_1$-node leaves us with the finite $A_5$-diagram or the removal of the $\alpha_2$-node yields a finite $D_5$-diagram. However, the decomposition rule for the hyperbolic algebras is not satisfied, as the removal of the  $\alpha_{-2}$-node produces a diagram of hyperbolic type.   

As indicated in the Dynkin diagram, by the use of $\circ$ and $\bullet$ for the vertices, the major difference compared to the  $\mathbf{(A}_{2}\mathbf{)}_{-2}$-case is that this version allows for a bicolouration. This fact can be exploited in a technical manner as will be seen below. In close analogy to the discussion in the previous section we choose a 4+4 dimensional standard representation for the  $\mathbf{(A}_{3}\mathbf{)}_{-2}$-roots $\alpha_i$, $i=-2,\cdots,3$, 
\begin{eqnarray}
	\alpha _{1} &=&\left( 1,-1,0,0;0,0,0,0 \right)
	,\quad \, \alpha _{2}=\left(0,1,-1,0;0,0,0,0 \right) , \quad
    \,\,\,\, \,\,\alpha _{3}=\left(0,0,1,-1,0;0,0,0,0 \right),  \notag	\\
	\alpha _{0}&=&\left(-1, 0, 0, 1; 1, 0, 0, 0 \right) , \quad
	\alpha _{-1} =\left(0, 0,0,0;-1,1,0,0 \right) ,\quad \alpha _{-2}=\left(
	0,0,0,0;1,0,-1,1\right) , \notag
\end{eqnarray}
with Lorentzian inner product 
\begin{equation}
	x\cdot y = x_1 y_1 + x_2 y_2 + x_3 y_3 + x_4 y_4 - x_5 y_6 - x_6 y_5- x_7 y_8 - x_8 y_7,
\end{equation}
for arbitrary 4+4 dimensional vectors $x=(x_1, \ldots x_8)$, $y=(y_1, \ldots y_8)$. Demanding the length of a generic root 
\begin{equation}
	\alpha = p \alpha_{-2} + q \alpha_{-1} + l \alpha_0 + m \alpha_1 + n \alpha_2 + r \alpha_3 \label{alpha}, \qquad   p,q,l,m,n,r \in \mathbb{Z},
\end{equation}
to be 2, leads now to the slightly modified Diophantine equation when compared to the previous sections
\begin{equation}
l^2-l m-l q-l r+m^2-m n+n^2-n r+p^2-p q+q^2+r^2 = 1 \quad \Leftrightarrow \quad \alpha \cdot \alpha =2 . \label{Dio2}
\end{equation}
The Weyl reflections are defined in the same manner as in equation (\ref{defWeyl}) with the additional map $\sigma_3$ associated to the root $\alpha_3$. The symmetries of the Diophantine equation (\ref{Dio2}) are now generated by the Weyl reflections as
\begin{eqnarray}
	&&\sigma_{-2}(\alpha) : p \rightarrow q-p , \quad \quad \,
	\sigma_{-1}(\alpha) : q \rightarrow l+p-q , \quad \,\,\,\,
	\sigma_{0}(\alpha) : l \rightarrow q+m+r-l , \quad  \label{Weylr12} \\
	&&\sigma_{1}(\alpha) : m \rightarrow l+n-m , \quad 
	\sigma_{2}(\alpha) : n \rightarrow m +r - n  , \quad
	\sigma_{3}(\alpha) : r \rightarrow l + n - r . \notag
\end{eqnarray}
Using the notations from \cite{Mass2} we can now use the bicolouration to define to specific elements consisting of commuting Weyl reflections whose mutual product makes up a Coxeter element 
\begin{equation}
	\sigma_+ := \sigma_{-1} \sigma_1 \sigma_3, \quad
	\sigma_- := \sigma_{-2} \sigma_0 \sigma_2, \quad
	\sigma_{L'} := \sigma_- \sigma_+. \quad
\end{equation}  
We verify now that the Kostant identity \cite{Kost} established originally for semi-simple algebras also holds for the $\mathbf{(A}_{3}\mathbf{)}_{-2}$-root space
\begin{equation}
	\left( \sigma_- + \sigma_+ \right) \alpha_i = \sum_{j=-2}^3 \left( 2 \delta_{ij} - K_{(3+i)(3+j)} \right) \alpha_j, \qquad i=-2, \cdots , 3.   \label{Kostantid}
\end{equation}   
Using this relation, together with $\sigma_{L'}^{-1}= \sigma_+ \sigma_-$ one readily derives a relation between the eigensystems of the Cartan matrix and the Coxeter element. Defining the vectors 
\begin{eqnarray}
	q_i^+ := \zeta_{i4} \alpha_1  + \zeta_{i6} \alpha_3  + \zeta_{i2} \alpha_{-1}, \quad \text{and} \quad
	q_i^- := \zeta_{i5} \alpha_2  + \zeta_{i3} \alpha_0  + \zeta_{i1} \alpha_{-2},
\end{eqnarray}  
where $\zeta = (v_1, \cdots , v_6)$ is the matrix with columns corresponding to the eigenvectors of the Cartan matrix
\begin{eqnarray}
K v_j = \lambda_j v_j,  \label{eigCart}
\end{eqnarray}  
we can explicitly define the eigenvectors of the Coxeter element as 
\begin{eqnarray}
q_j = e^{i \theta_j/2} q_j^- + e^{-i \theta_j/2} q_j^+ .
\end{eqnarray}  
The eigenvalue equation then reads
\begin{equation}
	\sigma q_j = e^{i 2 \theta_j} q_j ,  \label{EigCox}
\end{equation}
where the eigenvalues of the Cartan matrix in (\ref{eigCart}) and those of the Coxeter element in (\ref{EigCox}) are related as
\begin{equation}
	\lambda_j = 2 - 2\cos{\theta_j}  .  \label{Eigrel}
\end{equation}
We notice that there is no equivalent quantity corresponding to the integer exponents labelling angles $\theta_j$ in the finite dimensional case, however, they still add up pairwise to $\pi$ 
\begin{equation}
	\theta_j + \theta_{7-j} = \pi,  \qquad j=1, \cdots , 6.
\end{equation}
Concretely we have
\begin{equation}
	\theta_1= \arcsec \left(\frac{-2}{\sqrt{3+\sqrt{3}}}\right) = \pi - \theta_6, \quad
	\theta_2= \arcsec\left(-\sqrt{2+\frac{2}{\sqrt{3}}}\right) \pi - \theta_5, \quad
	\theta_3= \frac{\pi}{2}= \theta_4 . \quad
\end{equation}
 It is clear from these values that the order of the Coxeter element can not be finite, as we expect. Finite order $\sigma^h =1$ with $h \in \mathbb{N}$ would correspond $h \theta_i = n \pi$, with $n \in \mathbb{Z}$, for which there is no solution.  

\subsection{Invariants and integrability}
Let us now address the question of whether one can build solvable physical models that respect an $\mathbf{(A}_{3}\mathbf{)}_{-2}$-symmetry. In many model, such as those with Calogero type potentials, this requirement relates to the existence of invariants of the Weyl group. One may either use the invariants directly to construct a solvable model \cite{HR1,HR3,Fring:2004zw} or identify them within a given model to establish its integrability \cite{dunkl89,carrillo21}. 

Thus to each vector $x = \sum_{i=1}^6 x_i \alpha_{i-3} \in \mathbb{R}^8$ we associate a polynomials $P(x_1, \cdots , x_6)$ that remains invariant under the action of the Weyl group $\cal{W}$
\begin{equation}
	       P(\omega^{-1}(x))=P(x')=P(x)  \label{invdef}
\end{equation}
for all $\omega \in \cal{W}$ with $\omega^{-1}(x) \rightarrow x'$. The transformation of the components $x_i$ are simply obtained from (\ref{Weylr12}) with the replacements $p \rightarrow x_1, q \rightarrow x_2,l \rightarrow x_3, m \rightarrow x_4,n \rightarrow x_5,r \rightarrow x_6$. It is then easy to verify that there is no $\mathbf{(A}_{3}\mathbf{)}_{-2}$ polynomial invariant of order 1. At order 2 we find the standard invariant
\begin{equation}
	I_2(x_1, \cdots , x_6) = \sum_{i=1}^6 x_i^2 + \sum_{i,j=1}^6 K_{ij} x_i x_j . \label{Pinv}
\end{equation}  
To proceed to higher order one can make an Ansatz of a polynomial of a particular order with arbitrary coefficients and then fix them by systematically using (\ref{invdef}) for every generator of the Weyl group. While this is possible, especially with the help of symbolic algebra systems software, we will provide a different more direct option. For this purpose we carry out a change of variables from $x$ to $w$ defined in an implicit way by the relation
\begin{equation}
       x = \sum_{i=1}^6 x_i \alpha_{i-3} =  \sum_{i=1}^6 w_i q_i . \label{vartrans}
\end{equation} 
From relation (\ref{EigCox}) it is clear that the coordinates $w_i$ transform under the action of the Coxeter element as $\sigma: w_j \rightarrow e^{2 i \theta_j} w_j $. This implies that in these new coordinates any invariant of order $n$ must be of the form
\begin{equation}
           I_n(w_1, \ldots , w_6) = \sum_{a_1, \ldots , a_6 =0}^n c(a_1, \ldots , a_6)  w_1^{a_1} \ldots w_6^{a_6},  \quad  \,\,\sum_{i=1}^6 a_i =n, \,\, \sum_{i=1}^6 a_i \theta_i = m \pi, \,m \in \mathbb{Z}  .  \label{ingenex}
\end{equation} 
Since by construction each term in the sum is invariant under the action of the Coxeter element the constant coefficients $c(a_1, \ldots , a_6)$ must be constrained further by utilising the Weyl reflections associated to the simple roots. Checking this for every generator of the Weyl group  we find  
\begin{equation}
	I_2(w_1, \ldots , w_6) = i \left[w_3^2-4 \left(3+2 \sqrt{3}\right) w_2 w_5 -4 \left(3-2 \sqrt{3}\right) w_1 w_6\right] . \label{i2w}
\end{equation} 
Alternatively we also obtain (\ref{i2w}) with the variable transformation  $x_i \rightarrow w_i$ of the invariant (\ref{Pinv}), as specified in (\ref{vartrans}). It is now easy to verify that for $n$ odd we can never solve the last two equation in (\ref{ingenex}). Moreover, for $n=4$ we verified for the general Ansatz  (\ref{ingenex}) that the only invariant at that level is $	I_4=	I_2^2$. This means we can not build an exactly solvable $\mathbf{(A}_{3}\mathbf{)}_{-2}$ invariant model of Calogero type. In turn this suggest that the $\mathbf{(A}_{3}\mathbf{)}_{-2}$ symmetric models proposed in the next section are likely to be not integrable.

\section{Calogero type models associated to Lorentzian Kac–Moody algebras}	
We return now to our $\mathbf{(A}_{2}\mathbf{)}_{-2}$ example and start the discussion with the kinetic energy term for which we also need to use the inner product as defined in equation (\ref{definner}), so that it reads
\begin{equation}
	H_{\text{kin}} = \frac{1}{2} p \cdot p = \frac{1}{2} \left(  p_1^2 + p_2^2 +p_3^2 - 2 p_4 p_5 - 2 p_6 p_7 \right)  . \label{ekin}
\end{equation}
Since Weyl reflections are orthogonal transformation, i.e. $p \cdot p = \sigma_i(p) \cdot \sigma_i(p)$, the kinetic term is by construction invariant under the actions of the $(A_2)_{-2}$-Weyl group. We will specify below how the Weyl reflections act on the momenta so that one may also verify this symmetry explicitly.    

Next we consider potentials of Calogero type by extending the notions from the well-studied cases related to finite dimensional semi-simple Lie algebras $\bf{g}$. All potentials considered here are of the general standard Calogero form
\begin{equation}
	V(q) = \sum_{\alpha \in \Delta_{\bf{g}} }  \frac{c_{\alpha}}{\left( \alpha \cdot q \right)^2 } = \sum_{\alpha \in \Delta_{\bf{g}} }  \frac{c_{\alpha}}{\left( \sigma_i (\alpha) \cdot q \right)^2 }
	=\sum_{\alpha \in \Delta_{\bf{g}} }  \frac{c_{\alpha}}{\left( \alpha \cdot \sigma_i (q) \right)^2 }, \quad i=-2,-1,0,1,2.  \label{Calo1}
\end{equation}
Inteerpreting now $\bf{g}$ a Lorentzian algebra, the potentials in (\ref{Calo1}) are by construction invariant under the action of the Lorentzian Weyl group as the sum extends over the entire group. In this case it extents to infinity. As indicated in (\ref{Calo1}), we may also realise the transformations on the roots in the dual coordinate space since the Weyl reflections are orthogonal transformation, $\left( \sigma_i (\alpha) \cdot q \right) = \left( \sigma_i^2 (\alpha) \cdot \sigma_i(q) \right) = \left( \alpha\cdot \sigma_i (q) \right) $. For instance, for our $(A_2)_{-2}$-Weyl group in the seven dimensional representation the actions (\ref{LWeylref}) may be realised in the coordinate space as
\begin{eqnarray}
\sigma_{-2}(q) 	&:& q_4 \rightarrow q_4+q_5+q_6 -q_7 ,  \quad q_6 \rightarrow -q_5-q_6, \quad  q_7 \rightarrow q_5-q_7, \label{WW1} \\ 
	\sigma_{-1}(q) &:& q_4 \rightarrow q_5 ,  \quad q_5 \rightarrow q_4 ,\\
	\sigma_{0}(q) &:& q_1 \rightarrow q_3-q_5 ,  \quad q_3 \rightarrow q_1+q_5, \quad  q_4 \rightarrow q_1-q_3+q_4 + q_5, \\ 
	\sigma_{1}(q) &:& q_1 \rightarrow q_2 ,  \quad q_2 \rightarrow q_1, \\
	\sigma_{2}(q) &:& q_2 \rightarrow q_3 ,  \quad q_3 \rightarrow q_2. \label{WW5}
\end{eqnarray}
The affine (\ref{affCox}), the hyperbolic (\ref{hypCox}) and the Lorentzian Coxeter elements (\ref{CoxLor}) therefore act on the coordinates as
\begin{eqnarray}
	\sigma_a(q) &:& q_1 \rightarrow q_2 ,  \quad q_2 \rightarrow q_1+q_5, \quad  q_3 \rightarrow q_3-q_5, \quad q_4 \rightarrow q_1-q_3+q_4+q_5,  \label{Coxaff1} \\
	\sigma_h(q) &:& q_1 \rightarrow q_2 ,  \quad q_2 \rightarrow q_1+q_4, \quad  q_3 \rightarrow q_3-q_4,  \quad
	 q_4 \rightarrow q_1-q_3+q_4- q_5  ,  \quad q_5 \rightarrow q_4, \\
	\sigma_L(q) &:& q_1 \rightarrow q_2 ,  \quad q_2 \rightarrow q_1+q_4+q_5+q_6 -q_7, \quad  q_3 \rightarrow q_3-q_4-q_5-q_6 +q_7, \quad \\ &&  q_4 \rightarrow q_1-q_3+q_4- 2 q_5 + q_6 -q_7 ,  \quad q_5 \rightarrow q_4+q_5+q_6 -q_7 ,  \quad q_6 \rightarrow -q_5-q_6, \notag \quad \quad \\ && q_7 \rightarrow q_5-q_7, \notag
\end{eqnarray}
respectively. The action on the momenta is taken to be the same. Thus using (\ref{WW1})-(\ref{WW5}) with $q \rightarrow p$, we may now verify explicitly the invariance of the kinetic energy term (\ref{ekin}) under the action of the $(A_2)_{-2}$-Weyl group.  

The two variants of the standard Calogero potential in (\ref{stanCalo1}) are now generalised in a straightforward fashion. Taking $\gamma_i$ to be the, say $\ell$, representatives of each orbit the potential in (\ref{Calo1}) can be expressed in the equivalent form
 \begin{equation}
 	V_C(q) = \sum_{i=1}^\ell  \sum_{n=-\infty}^{\infty}  \frac{ g_{in}}{\left[ \sigma^n ( \gamma_i) \cdot q \right]^2 } ,  \label{Calo2}
\end{equation}
where the $g_{in}$ are some new real coupling constants related to the $c_{\alpha}$. It will remain an open question whether $\ell $ can be taken to be finite in general. The generalisation of the first variant in (\ref{stanCalo1}) may be written more explicitly as  
 \begin{equation}
	V_D(q) =   \!\!\!\!\!\!\!  \sum_{\substack{p,q,l,m,n=0 \\ \text{Diophantine equn}} }^\infty  \!\!\!\!\!   \frac{g_{pqlmn}}{[ (p \alpha_{-2}+ q \alpha_{-1} + l \alpha_{0} + m \alpha_{1} + n \alpha_{2} )   \cdot q]^2}  \label{summ2} 
\end{equation}
 When restricting to particular levels of the Lie algebraic representation as suggested in \cite{lechtenfeld2022} for a hyperbolic case, one may achieve to carry out some, or possibly all of the infinite sums, at that level. However, the potentials obtained in this manner are not invariant under the infinite Weyl group. In contrast, aiming to achieve the latter we explore here also the version in equation (\ref{Calo2}). 

\subsection{$(A_2)_{-2}$-Calogero type potentials}	

 We will now investigate the potentials in more detail and discuss under which circumstances the infinite sums may be carried out explicitly.
 
\subsubsection{Invariant potentials of the affine Weyl group}	
Thus let us now construct some invariant potentials from the version (\ref{Calo2}), which has the advantage that it only involves one infinite sum, as in this case $\ell$ will be shown to be finite. Choosing as an example $\gamma_1 = \alpha_2$ as a representative of an orbit and taking all coupling constants to be the same, i.e. $g_{in}= g$, we may compute the particular term
 \begin{equation}
	V_1(q) = \sum_{n=-\infty}^{\infty}  \frac{ g}{\left[ \sigma_a^n ( \alpha_2) \cdot q \right]^2 } .
\end{equation}  
Using our general formula (\ref{Coxeteraffgen}) for the action of any power of the Coxeter element action on any $(A_2)_{-2}$-root we obtain 
\begin{equation}
	V_1(q) = \sum_{n=-\infty}^{\infty} \frac{16 g}{\left\{2 \left[(-1)^n-1\right] q_1-2 \left[(-1)^n+1\right] q_2+\left[-6 n+(-1)^n-1\right] q_5+4 q_3\right\}^2}, 
\end{equation}
when acting with $\sigma_a$ on $\alpha_2$ in the denominator. Splitting the sum into its even and odd part, i.e. $\sum_n=\sum_{2n} + \sum_{2n-1}$, they are easily evaluated separately when using the general formula $ \sum_{n=-\infty}^{\infty} (A+B n)^{-2} = \pi^2/ [\sin^2(A \pi /B) B^2]    $. In this way we obtain
\begin{equation}
	V_1(q) = \frac{\pi^2}{9 q_5^2} \left\{  \frac{g}{\sin^2\left[ \frac{\pi}{3 q_5} (q_2- q_3)  \right]   }  
	+  \frac{g}{\sin^2\left[ \frac{\pi}{3 q_5} (q_1- q_3-q_5)  \right]   }  \label{pot111}
	 \right\}  .
\end{equation} 
By construction this term is invariant under the action of the affine Coxeter element when realised in the coordinate space by (\ref{Coxaff1}). This term is, however,  not yet invariant under the entire Weyl group as for this we require more representatives of the orbits. We may either try with different roots for the $\gamma_i$ representatives, which equivalently corresponds to actions of the Weyl group on the coordinates as specified in (\ref{WW1})-(\ref{WW5}). Let us therefore consider the potential 
\begin{eqnarray}
	V(q) &=&  \sum_{n=-\infty}^{\infty}  \frac{ g}{\left[ \sigma^n ( \alpha_2) \cdot q \right]^2 } +
	 \frac{ g}{\left[ \sigma^n ( \alpha_2) \cdot \sigma_0(q) \right]^2 }
	 + \frac{ g}{\left[ \sigma^n ( \alpha_2) \cdot \sigma_1(q) \right]^2 }  \label{invWeylpot}\\
	&& + \frac{ g}{\left[ \sigma^n ( \alpha_2) \cdot \sigma_1\sigma_0(q) \right]^2 } 
	+ \frac{ g}{\left[ \sigma^n ( \alpha_2) \cdot \sigma_0\sigma_1(q) \right]^2 }
	  + \frac{ g}{\left[ \sigma^n ( \alpha_2) \cdot \sigma_0\sigma_1\sigma_0(q) \right]^2 } \notag \\
	 && + \frac{ g}{\left[ \sigma^n ( \alpha_2) \cdot \sigma_2\sigma_0(q) \right]^2}  
	 + \frac{ g}{\left[ \sigma^n ( \alpha_2) \cdot \sigma_2\sigma_1(q) \right]^2 }
	 + \frac{ g}{\left[ \sigma^n ( \alpha_2) \cdot \sigma_2\sigma_0\sigma_2(q) \right]^2 } . \notag
\end{eqnarray}
At this point the potential appears in an ad hoc manner, but we shall explain below how is systematically obtained. Crucially, we can evaluate each infinite sum as outlined above for the first term.  In this manner we obtain the potential
\begin{equation}
	V(q) = \frac{2\pi^2 g}{9 q_5^2} \left( V_{12}+ V_{13}+  V_{23} + V_{125}^+ + V_{125}^- + V_{135}^+  + V_{135}^- + V_{235}^+ +  V_{235}^-    \right) , \label{affinvpot}
\end{equation} 
where we abbreviated
\begin{equation}
	V_{ij}:= \frac{1}{ \sin^2 \left[ \frac{\pi}{3 q_5} (q_i- q_j)  \right] }, \quad
	V_{ijk}^\pm:= \frac{1}{\sin^2\left[ \frac{\pi}{3 q_5} (q_i- q_j \pm q_k)  \right]},   \quad   i,j,k= 1,2,3,4,5.	  
\end{equation}
We verify explicitly that this potential is indeed invariant under the affine Weyl group. Indicating how each term transforms under the action of the three generators we have
\begin{equation}
\begin{tabular}{ l  l l l l l }
	$\sigma_0: $&$ V_{12} \leftrightarrow V_{235}^+  $ \,\,&$ V_{13} \leftrightarrow V_{135}^-  $\,\, &$  V_{23} \leftrightarrow V_{125}^+ $ \,\, &$  V_{125}^- \leftrightarrow V_{235}^-$ \,\,
	& $  V_{135}^+ \circlearrowleft $ ,  \\ 
	$\sigma_1: $&$ V_{13} \leftrightarrow V_{23}  $ \,\,&$ V_{125}^+ \leftrightarrow V_{125}^-  $\,\, &$  V_{135}^- \leftrightarrow V_{235}^- $ \,\, &$  V_{135}^+ \leftrightarrow V_{235}^+$ \,\,
	& $  V_{12} \circlearrowleft $ , \\
	$\sigma_2: $&$ V_{12} \leftrightarrow V_{13}  $ \,\,&$ V_{125}^+ \leftrightarrow V_{135}^+  $\,\, &$  V_{235}^+ \leftrightarrow V_{235}^- $ \,\, &$  V_{125}^- \leftrightarrow V_{135}^-$ \,\,
	& $  V_{23} \circlearrowleft $ .
\end{tabular} \label{sigV}
\end{equation}
The action of the Coxeter element is therefore
\begin{equation}
	\begin{tabular}{ l  l l l l l }
		$\sigma_a: $&$ V_{12} \leftrightarrow V_{125}^+  $ \,\,&$ V_{13} \leftrightarrow V_{235}^+  $\,\, &$  V_{23} \leftrightarrow V_{135}^- $ \,\, &$  V_{235}^- \leftrightarrow V_{135}^+$ \,\,
		& $  V_{125}^- \circlearrowleft $  .
	\end{tabular}
\end{equation}
Thus the potential (\ref{affinvpot}) is invariant under the entire $(A_2)_{0}$-affine Weyl group. It is now also more obvious how the potential (\ref{invWeylpot}) was obtained in the first place. Starting off with the terms in $V_1$, we may utilise the Weyl symmetries (\ref{sigV}) to generate additional terms until the symmetries only generate terms already included. As it turned out nine terms are sufficient to achieve that.

Furthermore we notice that we recover the standard $A_2$-Calogero model in the infinite limits
\begin{equation}
	\lim_{q_5 \rightarrow \pm \infty} V(q) = 2 g \left[  \frac{1}{(x_1 - x_2)^2 } +  \frac{1}{(x_1 - x_3)^2 }  + \frac{1}{(x_2 - x_3)^2 }       \right] .
\end{equation}

Let us now compare this potential with the one obtained when evaluating the three infinite sums in (\ref{summ2}) constrained by the Diophantine equation (\ref{Dioph}) with $p=q=0$. For this purpose we re-write (\ref{summ2}) and (\ref{Calo2}) as
\begin{equation}
	V_D(q)  =    g  \!\!\!\!\!\!\!  \!\!\!\!\!\!\!   \sum_{\substack{l,m,n=0 \\ \text{Diophantine equn}} }^\infty    \!\!\!\!\!\!\!   \!\!\!\!\!\!\!  v_{lmn}, \qquad \text{and} \qquad
	V_C(q) =  g \sum_{n=-\infty}^{\infty} \sum_{i=1}^9  v_i(n)   ,  \label{summ33}
\end{equation}
respectively, where the $v_i(n) $ are identified in the same order as the terms appears in (\ref{Calo2}). We have set all coupling constants to be equal. Defining the quantities
\begin{equation}
	v_{ij}= \frac{1}{(q_i-q_j)^2},  \qquad \text{and} \qquad 
		v_{ijk}^\pm(n)= \frac{1}{(q_i-q_j \pm n q_k)^2},
\end{equation}
we list all nonzero terms in (\ref{summ33}) with the upper limit in $V_D(q)$ taken to be 5. We find
\begin{equation}
	\begin{tabular}{ l l l l}
		$v_{001}= v_1(0)= v_{23}$,  & $v_{010}= v_8(0)= v_{12}$,  & $v_{011}= v_3(0)= v_{13}$,  \\   
		$v_{100}= v_7(0)= v^+_{135}(1)$,  &$v_{101}= v_2(0)= v^+_{125}(1)$,  & $v_{110}= v_9(1)= v^+_{235}(1)$, \\ 
		 $v_{112}= v_3(-1)= v^-_{235}(1)$, &  $v_{121}= v_4(0)= v^-_{125}(1)$,  &   $v_{122}= v_9(0)= v^-_{135}(1)$, \\
		  $v_{211}= v_5(0)= v^+_{135}(2)$, &  $v_{212}= v_5(-1)= v^+_{125}(2)$,  &   $v_{221}= v_6(0)= v^+_{235}(2)$, \\   	  
		   $v_{223}= v_9(-1)= v^-_{235}(2)$, &  $v_{232}= v_6(-1)= v^-_{125}(2)$,  &   $v_{233}= v_7(2)= v^-_{235}(2)$, \\ 
		   $v_{322}= v_3(2)= v^+_{135}(3)$, &  $v_{323}= v_8(2)= v^+_{125}(3)$,  &   $v_{332}= v_1(2)= v^+_{235}(3)$, \\ 
		   $v_{334}= v_1(-2)= v^-_{235}(3)$, &  $v_{343}= v_8(-2)= v^-_{125}(3)$,  &   $v_{344}= v_3(-2)= v^-_{135}(3)$, \\ 
		   $v_{433}= v_9(2)= v^+_{135}(4)$, &  $v_{434}= v_2(-2)= v^+_{125}(4)$,  &   $v_{443}= v_9(3)= v^+_{235}(4)$, \\ 
		   $v_{445}= v_3(-3)= v^-_{235}(4)$, &  $v_{454}= v_5(3)= v^-_{125}(4)$,  &   $v_{455}= v_1(-3)= v^-_{135}(4)$, \\    
		    $v_{544}= v_1(3)= v^+_{135}(5)$, &  $v_{545}= v_5(-3)= v^+_{125}(5)$,  &   $v_{554}= v_3(3)= v^+_{235}(5)$.
		\end{tabular}  \label{firstterms}
\end{equation}
One may have suspected that $V_C(q)$ is only a subset of the general expression $V_D(q)$. However, we observe that the potential in the form of summing over Coxeter orbits, $V_C(q)$ captures all the terms from the version expressed in terms of three infinite sums subject to the Diophantine equation $V_D(q)$. We have verified this beyond the order reported here. This constitutes of course no rigorous proof of the equality of the two versions, but very strong evidence.	

\subsubsection{Invariant potentials of the hyperbolic Weyl group}	
Next we analyse the two versions of the general potential by just keeping  $p=0$ in (\ref{summ2}) and by taking  in (\ref{Calo2}) $\sigma=\sigma_h$ to be the Coxeter element related to the hyperbolic Weyl group. Consequently, the resulting potentials will be invariant under the action of the entire hyperbolic Weyl group. For this purpose we re-write (\ref{summ2}) and (\ref{Calo2}) as 
\begin{equation}
	V_D(q)  =    g  \!\!\!  \!\!\!   \!\!\! \sum_{\substack{q,l,m,n=0 \\ \text{Diophantine equn}} }^\infty   \!\!\!   \!\!\!  \!\!\!\!  v_{qlmn}, \qquad \text{and} \qquad
	V_C(q) =  g \sum_{k=-\infty}^{\infty} \gamma_i^k  ,  \label{summ3}
\end{equation}
respectively. Once again we set all the coupling constants to be equal. We use the abbreviations
\begin{equation}
	\gamma_i^k  := \frac{1}{ [ \sigma_h^k( \gamma_i ) \cdot q  ]^2   }, \quad \text{and} \quad w_{r,s,t,u,v}:= \frac{1}{(r q_1 + s q_2 + t q_3 + u q_4 + v q_5 )^2}.
\end{equation}
In version (\ref{Calo2}) we need to generate sufficiently many representatives $\gamma_i$ of the respective orbits. 
We list now all the terms appearing in the potential $V_D(q)$ up to the order $5$ and identify an equivalent way to express them as terms appearing in the version $V_C(q) $. When $q=0$ all the terms in (\ref{firstterms}) also appear, but as we now employ a different type of Coxeter element they are differently realised in formula $V_C(q)$. We find
\begin{equation}
	\begin{tabular}{ l l l l}
		$v_{0001} = v_{23}= \gamma_1$,  \qquad \qquad \qquad & $v_{0010}= v_{12}=  \gamma_2 $,  \qquad \qquad \qquad & $v_{0011}= v_{13}=  \gamma_3 $,  \\   
		$v_{0100} = v^+_{135}(1)= \gamma_4$,  &$v_{0101} = v^+_{125}(1)=  \gamma_2^{-1}$,  & $v_{0110} = v^+_{235}(1)= \gamma_2^2$, \\ 
		$v_{0112} = v^-_{235}(1)=  \gamma_5$, &  $v_{0121} = v^-_{125}(1)= \gamma_6$,  &   $v_{0122}= v^-_{135}(1)=   \gamma_1^{-1}$, \\
		$v_{0211} = v^+_{135}(2)=  \gamma_7$, &  $v_{0212} = v^+_{125}(2)=  \gamma_8$,  &   $v_{0221} = v^+_{235}(2)=  \gamma_9$, \\   				  
		$v_{0223} = v^-_{235}(2)=  \gamma_3^{-1}$, &  $v_{0232}= v^-_{125}(2)=  \gamma_{10}$,  &   $v_{0233} = v^-_{235}(2)=  \gamma_4^{-3}$, \\ 
		$v_{0322} = v^+_{135}(3)=  \gamma_{11}$, &  $v_{0323} = v^+_{125}(3)=  \gamma_{12}$,  &   $v_{0332} = v^+_{235}(3)=  \gamma_{13}$, \\ 
		$v_{0334} = v^-_{235}(3)=  \gamma_1^{-3}$, &  $v_{0343} = v^-_{125}(3)=  \gamma_{14}$,  &   $v_{0344} = v^-_{135}(3)= \gamma_7^{-2} $, \\ 
		$v_{0433} = v^+_{135}(4)=  \gamma_{15}$, &  $v_{0434} = v^+_{125}(4)=  \gamma_{16}$,  &   $v_{0443} = v^+_{235}(4)=  \gamma_{17}$, \\ 
		$v_{0445} = v^-_{235}(4)=  \gamma_{18}$, &  $v_{0454} = v^-_{125}(4)=  \gamma_{19}$,  &   $v_{0455}= v^-_{135}(4)= \gamma_{20} $, \\    
		$v_{0544} = v^+_{135}(5)=  \gamma_{21}$, &  $v_{0545}= v^+_{125}(5)=   \gamma_{22}$,  &   $v_{0554} = v^+_{235}(5)= \gamma_{23}$.
	\end{tabular}  
\end{equation}
Extending the sum in $p$ beyond $0$, we find in addition
\begin{equation}
	\begin{tabular}{ l l l l}
$v_{1000}=w_{0,0,0,1,-1}=\gamma_4^{-1}$,   &   $v_{1100}  =  w_{1,0,-1,1,0}=\gamma_5^1$,  &   $v_{1101} = w_{1,-1,0,1,0}=\gamma_2^1$,  \\
$v_{1110}=w_{0,1,-1,1,0}=\gamma_3^1$,   &  $v_{1112}=w_{0,1,-1,-1,0}= \gamma_4^{-2}$,  &   $v_{1121}=w_{1,-1,0,-1,0}= \gamma_8^1$, \\
$v_{1122}=w_{1,0,-1,-1,0}=\gamma_2^{-2}$,   &    $  v_{1312}=w_{2,-1,-1,1,2} =  \gamma_{18}^2$,      &    $ v_{1321}=w_{1,1,-2,1,2} =  \gamma_6^2 $, \\
$  v_{1324}=w_{1,-2,1,1,2} =  \gamma_6^{-1}$,    &    $  v_{1342}=w_{1,-2,1,-1,-2} = \gamma_{24} $,    &   $ v_{1345}=w_{1,1,-2,-1,-2}=\gamma_5^{-1}  $, \\   
  $  v_{1354}=w_{2,-1,-1,-1,-2} = \gamma_{25}$,    &   $  v_{1422}=w_{2,0,-2,1,3}= \gamma_{20}^2$ ,   &   $ v_{1424}=w_{2,-2,0,1,3} = \gamma_{26}$,  
       \end{tabular} 
      \end{equation}
      \begin{equation}  
    \begin{tabular}{ l l l l}
    	$ v_{1442}=w_{0,2,-2,1,3} = \gamma_{27}$,    &   $ v_{2211}=w_{1,0,-1,2,0} = \gamma_1^1$,    &   $   v_{2212}=w_{1,-1,0,2,0}= \gamma_6^1$ , \\
    	$  v_{2221}=w_{0,1,-1,2,0} =\gamma_4^1 $,    &   $  v_{2223}=w_{0,1,-1,-2,0} =\gamma_7^{-1}$,    &   $ v_{2232}=w_{1,-1,0,-2,0}= \gamma_{12}^1$,  \\  
    	$  v_{2233}=w_{1,0,-1,-2,0} = \gamma_{18}^1$,    &    $ v_{2312}=w_{2,-1,-1,2,1} = \gamma_8^2$,    &   $ v_{2321}=w_{1,1,-2,2,1} = \gamma_5^2$,  \\  
    	 $  v_{2324}=w_{1,-2,1,2,1} = \gamma_9^{-1}$,    &    $ v_{2342}=w_{1,-2,1,-2,-1}= \gamma_{26}^1  $,    &   $ v_{2345}=w_{1,1,-2,-2,-1}=  \gamma_8^{-1}$,  \\  
    	$   v_{2354}=w_{2,-1,-1,-2,-1} = \gamma_{28}$,    &    $ v_{2523}=w_{3,-1,-2,2,3} = \gamma_{29}$,    &   $ v_{2524}=w_{3,-2,-1,2,3} = \gamma_{30}$ , \\     
    	$ v_{2532}=w_{2,1,-3,2,3} =  \gamma_{10}^2$ ,   &  $  v_{2542}=w_{1,2,-3,2,3}= \gamma_{25}^2 $,    &   $ v_{3322}=w_{1,0,-1,3,0} =\gamma_2^3$, \\
    	$ v_{3323}=w_{1,-1,0,3,0} = \gamma_{10}^1 $,    &   $  v_{3332}=w_{0,1,-1,3,0} = \gamma_7^1$,    &   $ v_{3334}=w_{0,1,-1,-3,0} = \gamma_{20}^1$,  \\     
    	$  v_{3343}=w_{1,-1,0,-3,0} = \gamma_{16}^1$,    &    $  v_{3344}=w_{1,0,-1,-3,0} = \gamma_{29}^{-1}$,    &   $ v_{3422}=w_{2,0,-2,3,1}= \gamma_3^2 $,  \\  
    	$ v_{3424}=w_{2,-2,0,3,1}  = \gamma_{25}^1$,    &    $ v_{3442}=w_{0,2,-2,3,1}=  \gamma_{18}^3$,    &   $ v_{3523}=w_{3,-1,-2,3,2}= \gamma_{12}^2$,  \\
      $ v_{3524}=w_{3,-2,-1,3,2} = \gamma_{28}^1 $,    &    $ v_{3532}=w_{2,1,-3,3,2} = \gamma_1^2 $,    &   $ v_{3542}=w_{1,2,-3,3,2} = \gamma_8^3$,  \\    
  $  v_{4433}=w_{1,0,-1,4,0} = \gamma_9^1$,    &    $ v_{4434}=w_{1,-1,0,4,0} =  \gamma_{14}^1 $,    &   $ v_{4443}=w_{0,1,-1,4,0} = \gamma_{11}^1$,  \\ 
   $  v_{4445}=w_{0,1,-1,-4,0} = \gamma_{31}$ ,   &    $  v_{4454}=w_{1,-1,0,-4,0} = \gamma_{22}^1$,    &   $ v_{4455}=w_{1,0,-1,-4,0} = \gamma_{32}$,  \\
   $ v_{5544}=w_{1,0,-1,5,0}  = \gamma_{13}^1$,    &   $  v_{5545}=w_{1,-1,0,5,0} = \gamma_{19}^1$,    &   $ v_{5554}=w_{0,1,-1,5,0}  = \gamma_{15}^1$.  
\end{tabular} 
\end{equation}
We notice here that each term in the multiple infinite sums of the ``raw version" $V_D(q)$ has a counterpart in the version $V_C(q)$ that is based on orbits of the Coxeter element. We also observe that we require many more orbits, e.g. 35 up to level 5, as in the affine case to cover the entire root space. In this case it remains unresolved whether a systematic way exist to generate these orbits and whether a finite number of orbits will be sufficient to generate all terms of the potential. Note here that generating all terms in the potential is not equivalent to generating the entire Weyl group, as in the former we may exploit additional symmetries coming from the nature of the potential. 

\subsubsection{Invariant potentials of the Lorentzian Weyl group}	
Finally we compute a potential that is invariant under the full Lorentzian Weyl group. For this we generalise (\ref{summ3}) in an obvious manner summing now also over $p$. We require now the abbreviations 
\begin{equation}
	\gamma_i^k  := \frac{1}{ [ \sigma_L^k( \gamma_i ) \cdot q  ]^2   }, \quad \text{and} \quad w_{r,s,t,u,v,x,y}:= \frac{1}{(r q_1 + s q_2 + t q_3 + u q_4 + v q_5+ x q_6 + y q_7 )^2}.
\end{equation}
We list all the terms in $V_D(q)$ up to the order $3$ and identify a corresponding term in the potential $V_C(q) $
\begin{equation}
	{\small \begin{tabular}{ l l l}
			$\!\! v_{00001}=w_{0,1,-1,0,0,0,0}=\gamma_1 $, & $ \!\! \!\! v_{00010}=w_{1,-1,0,0,0,0,0} =\gamma _2$, & $ \!\! \!\! v_{00011}=w_{1,0,-1,0,0,0,0} =\gamma _{3} $, \\
			$\!\! v_{00100}=w_{1,0,-1,0,1,0,0} =\gamma _{4} $, & $ \!\! \!\! v_{00101} = w_{1,-1,0,0,1,0,0} =\gamma _2^{-1} $, & $ \!\! \!\! v_{00110} = w_{0,1,-1,0,1,0,0} =\gamma _5 $, \\
			$\!\! v_{00112} = w_ {0,1,-1,0,-1,0,0} =\gamma _6$, & $ \!\! \!\! v_{00121} = w_{1,-1,0,0,-1,0,0} =\gamma _7 $, & $ \!\! \!\! v_{00122} = w_{1,0,-1,0,-1,0,0} =\gamma _1^{-1} $, \\
			$\!\! v_{00211} = w_{1,0,-1,0,2,0,0} =\gamma _{8} $, & $ \!\! \!\! v_{00212} = w_{1,-1,0,0,2,0,0} =\gamma _{9} $, & $ \!\! \!\! v_{00221} = w_ {0,1,-1,0,2,0,0} =\gamma _{10}$, \\
			$\!\! v_{00223} = w_{0,1,-1,0,-2,0,0} =\gamma _{3}^{-1}$, & $ \!\! \!\! v_{00232} = w_{1,-1,0,0,-2,0,0} =\gamma _{11} $, & $ \!\! \!\! v_{00233} = w_{1,0,-1,0,-2,0,0} =\gamma _{4}^{-4}$, \\
			$\!\! v_{00322} = w_{1,0,-1,0,3,0,0} =\gamma _{12} $, & $ \!\! \!\! v_{00323} = w_ {1,-1,0,0,3,0,0} =\gamma _{13} $, & $ \!\! \!\! v_{00332} = w_{0,1,-1,0,3,0,0} =\gamma _{14} $, \\
			$\!\! v_{01000} = w_{0,0,0,1,-1,0,0} =\gamma _{4}^{-1} $, & $ \!\! \!\! v_{01100} = w_{1,0,-1,1,0,0,0} =\gamma _{15} $, & $ \!\! \!\! v_{01101} = w_{1,-1,0,1,0,0,0} =\gamma _5^{-1} $, \\
			$\!\! v_{01110} = w_{0,1,-1,1,0,0,0} =\gamma _2^2 $, & $ \!\! \!\! v_{01112} = w_{0,1,-1,-1,0,0,0} =\gamma _{15}^{-2} $, & $ \!\! \!\! v_{01121} = w_{1,-1,0,-1,0,0,0} =\gamma _{16} $, \\
			$\!\! v_{01122} = w_{1,0,-1,-1,0,0,0} =\gamma _2^{-2} $, & $ \!\! \!\! v_{01312} = w_{2,-1,-1,1,2,0,0} =\gamma _{17} $, & $ \!\! \!\! v_{01321} = w_{1,1,-2,1,2,0,0} =\gamma _{18} $, \\
			$\!\! v_{02211} = w_{1,0,-1,2,0,0,0} =\gamma _{19} $, & $ \!\! \!\! v_{02212} = w_{1,-1,0,2,0,0,0} =\gamma _{20} $, & $ \!\! \!\! v_{02221} = w_{0,1,-1,2,0,0,0} =\gamma _{21} $, \\
			$\!\! v_{02223} = w_{0,1,-1,-2,0,0,0} =\gamma _{8}^{-1}$, & $ \!\! \!\! v_{02232} = w_{1,-1,0,-2,0,0,0} =\gamma _{22} $, & $ \!\! \!\! v_ {02233} = w_{1,0,-1,-2,0,0,0} =\gamma _{23} $, 
			 \end{tabular} }
		\end{equation}
	\begin{equation}  
  {\small 	\begin{tabular}{ l l l l}
  		$\!\! v_{02312} = w_{2,-1,-1,2,1,0,0} =\gamma _{24}$, & $ \!\! \!\! v_{02321} = w_{1,1,-2,2,1,0,0} =\gamma _{25} $, & $ \!\! \!\! v_{03322} = w_{1,0,-1,3,0,0,0} =\gamma _{26} $, \\
			$\!\! v_{03323} = w_{1,-1,0,3,0,0,0} =\gamma _{27} $, & $ \!\! \!\! v_{03332} = w_{0,1,-1,3,0,0,0} =\gamma _{28} $, & $ \!\! \!\! v_{10000} = w_{0,0,0,0,1,1,-1} =\gamma _{4}^{-2} $, \\
			$\!\! v_{10111} = w_{0,0,0,0,2,1,-1} =\gamma _{29} $, & $ \!\! \!\! v_{10222} = w_{0,0,0,0,3,1,-1} =\gamma _{30} $, & $ \!\! \!\! v_{10333} = w_{0,0,0,0,4,1,-1} =\gamma _{31} $, \\
			$\!\! v_{11000} = w_{0,0,0,1,0,1,-1} =\gamma _{15}^{-1} $, & $ \!\! \!\! v_{11100} = w_{1,0,-1,1,1,1,-1} =\gamma _6^1 $, & $ \!\! \!\! v_{11101} = w_{1,-1,0,1,1,1,-1} =\gamma _2^1 $, \\
			$\!\! v_{11110} = w_{0,1,-1,1,1,1,-1} =\gamma _{3}^1 $, & $ \!\! \!\! v_{11112} = w_{0,-1,1,1,1,1,-1} =\gamma _{4}^{-3} $, & $ \!\! \!\! v_{11121} = w_{-1,1,0,1,1,1,-1} =\gamma _{9}^1 $, \\
			$\!\! v_{11122} = w_{-1,0,1,1,1,1,-1} =\gamma _5^{-2} $, & $ \!\! \!\! v_{11312} = w_{2,-1,-1,1,3,1,-1} =\gamma _{23}^1 $, & $ \!\! \!\! v_{11321} = w_{1,1,-2,1,3,1,-1} =\gamma _{20}^1 $, \\
			$\!\! v_{12111} = w_{0,0,0,2,0,1,-1} =\gamma _{29}^1 $, & $ \!\! \!\! v_{12311} = w_{2,0,-2,2,2,1,-1} =\gamma _{29}^2 $, & $ \!\! \!\! v_{12313} = w_{2,-2,0,2,2,1,-1} =\gamma _{32} $, \\
			$\!\! v_{12331} = w_{0,2,-2,2,2,1,-1} =\gamma _{32}^1 $, & $ \!\! \!\! v_{13222} = w_{0,0,0,3,0,1,-1} =\gamma _{33} $, & $ \!\! \!\! v_{13312} = w_{2,-1,-1,3,1,1,-1} =\gamma _{34} $, \\
			$\!\! v_{13321} = w_{1,1,-2,3,1,1,-1} =\gamma _7^2 $, & $ \!\! \!\! v_{21222} = w_{0,0,0,1,3,2,-2} =\gamma _{35}  $, & $ \!\! \!\! v_{21322} = w_{1,0,-1,1,4,2,-2} =\gamma _{36} $, \\
			$\!\! v_{21323} = w_{1,-1,0,1,4,2,-2} =\gamma _{37}  $, & $ \!\! \!\! v_{21332} = w_{0,1,-1,1,4,2,-2} =\gamma _{38} $, & $ \!\! \!\! v_{22211} = w_{1,0,-1,2,2,2,-2} =\gamma _1^1$, \\
			$\!\! v_{22212} = w_{1,-1,0,2,2,2,-2} =\gamma _7^1 $, & $ \!\! \!\! v_{22221} = w_{0,1,-1,2,2,2,-2} =\gamma _{4}^1 $, & $ \!\! \!\! v_{22223} = w_{0,-1,1,2,2,2,-2} =\gamma _{19}^{-1} $, \\
			$\!\! v_{22232} = w_{-1,1,0,2,2,2,-2} =\gamma _{13}^1 $, & $ \!\! \!\! v_{22233} = w_{-1,0,1,2,2,2,-2} =\gamma _{34}^{-1} $, & $ \!\! \!\! v_{22312} = w_{2,-1,-1,2,3,2,-2} =\gamma _{16}^1 $, \\
			$\!\! v_{22321} = w_{1,1,-2,2,3,2,-2} =\gamma _{15}^1 $, & $ \!\! \!\! v_{23222} = w_{0,0,0,3,1,2,-2} =\gamma _{30}^1 $, & $ \!\! \!\! v_{23312} = w_{2,-1,-1,3,2,2,-2} =\gamma _{9}^2 $, \\
			$\!\! v_{23321} = w_{1,1,-2,3,2,2,-2} =\gamma _6^2 $, & $ \!\! \!\! v_{32333} = w_{0,0,0,2,4,3,-3} =\gamma _{39} $, & $ \!\! \!\! v_{33322} = w_{1,0,-1,3,3,3,-3} =\gamma _5^1 $, \\
			$\!\! v_{33323} = w_{1,-1,0,3,3,3,-3} =\gamma _{11}^1 $, & $ \!\! \!\! v_{33332} = w_ {0,1,-1,3,3,3,-3} =\gamma _{8}^1 $,. &
	\end{tabular} }
\end{equation}
Once more we see that each term possess at least one counter term in each version of the potential. We leave it as open issue to construct all orbits, to find a closed formula or even computing the sums.

\subsubsection{Partial sums}	

Alternatively, to converting the five infinity sums into one infinite sum over orbits of the Coxeter element, we may also consider partial sums for specific choices of some of the summation indices compatible with the Diophantine equation with one infinite sum remaining. We already encountered partial sums in obtaining potentials of the form (\ref{pot111}). While this term was not invariant under the entire Weyl group, it was by construction invariant under the action of the Coxeter element. We list here some further possibilities:

 For instance, the choice $  p \rightarrow l, q \rightarrow 1+l, m \rightarrow l, n \rightarrow l$ satisfies the Diophantine equation (\ref{Dioph}) and leaves one infinite sum in the general potential (\ref{summ2}) . Taking also all coupling constants to be the same $g$ and using the 3+4 dimensional representation for the roots $\alpha_i$ we are left with
\begin{equation}
	V(q) = \sum_{l=0}^\infty \frac{g }{ A + B l + \frac{B^2}{4A} l^2   } = \frac{4 A }{B^2} \Psi \left(\frac{2 A}{B}\right) ,   \label{infsumVconc}
\end{equation}
where $ \Psi(x) := (\ln \Gamma[x])^{''} $ is a polygamma function and $A = q_4^2-2 q_5 q_4+q_5^2$, $B= 2 q_4^2+2 q_6 q_4-2 q_7 q_4-2 q_5^2-2 q_5 q_6+2 q_5 q_7$. The sum in (\ref{infsumVconc}) has been carried out explicitly. Concretely we obtain the potential
\begin{equation}
	V(q) =  \frac{g}{\left(q_4+q_5+q_6-q_7\right)^2}  \Psi \left(\frac{q_4-q_5} {q_4+q_5+q_6-q_7}\right) .   \label{infsumres}
\end{equation}

For the choice $  p \rightarrow l, q \rightarrow l, m \rightarrow 1+l, n \rightarrow 1+l$ we obtain in the same manner
\begin{equation}
	V(q) =  \frac{g}{ \left(q_4+q_5+q_6-q_7\right)^2 }  \Psi \left(\frac{q_2-q_1}{q_4+q_5+q_6-q_7}\right) .   \label{infsumres}
\end{equation}

For the choice $  p \rightarrow k, q \rightarrow 2k,l \rightarrow 2 k+1, m \rightarrow k, n \rightarrow k$ then subsequent infinite sum over $k$ yields 
\begin{eqnarray}
	V_1(q) &=&  \frac{\left(q_1-q_3+q_5\right)^2}{\left[ \left(q_1-q_3+q_5\right) \left(q_1-q_3+2 q_4+q_5+q_6\right)+\left(-q_1+q_3-7 q_5\right)
		q_7\right]^2}\\
	&&\times \Psi \left[ \frac{\left(q_1-q_3+q_5\right){}^2}{\left(q_1-q_3+q_5\right) \left(q_1-q_3+2 q_4+q_5+q_6\right)
		+\left(-q_1+q_3-7 q_5\right) q_7} \right] .\notag \label{infsumres2}
\end{eqnarray}

Another interesting choice is $  p \rightarrow -1-k, q \rightarrow -2-2k, m \rightarrow -1-k, n \rightarrow -1 -k, l \rightarrow -2k-1$, which would mean we extend the sums in (\ref{summ2}) to the negative range. In the case we obtain the potential
\begin{eqnarray}
	V_2(q) &=&  \frac{\left(q_1-q_3+q_5\right)^2}{\left[ \left(q_1-q_3+q_5\right) \left(q_1-q_3+2 q_4+q_5+q_6\right)+\left(-q_1+q_3-7 q_5\right)
		q_7\right]^2}\\
	&&\times \Psi \left[1- \frac{\left(q_1-q_3+q_5\right){}^2}{\left(q_1-q_3+q_5\right) \left(q_1-q_3+2 q_4+q_5+q_6\right)
		+\left(-q_1+q_3-7 q_5\right) q_7} \right] .\notag \label{infsumres2}
\end{eqnarray} 
Using that fact that $\Psi(x) + \Psi(1-x) = \pi^2/ \sin(\pi x)^2$ the sum of the two potentials yields
\begin{eqnarray}
	V_1(q)+ V_2(q) &=&  \frac{ \pi^2 \left(q_1-q_3+q_5\right)^2}{\left[ \left(q_1-q_3+q_5\right) \left(q_1-q_3+2 q_4+q_5+q_6\right)+\left(-q_1+q_3-7 q_5\right)
		q_7\right]^2}\\
	&&\times \sin \left[ \pi \frac{\left(q_1-q_3+q_5\right){}^2}{\left(q_1-q_3+q_5\right) \left(q_1-q_3+2 q_4+q_5+q_6\right)
		+\left(-q_1+q_3-7 q_5\right) q_7} \right]^{-2} .\notag \label{infsumres2}
\end{eqnarray} 
Evidently, these potentials are not invariant under the action of the full Lorentzian Weyl group, but they involve no infinite sum and admit a partial symmetry.

\section{Conclusions}	
We discussed general properties of infinite Weyl groups of affine, hyperbolic and Lorentzian type.  As concrete examples we presented the $(A_{2})_0$, $(A_{2})_{-1}$ and $(A_2)_{-2}$ cases, for which we paid particular attention to the properties of the respective Coxeter elements. In each case we derived closed formulae for the infinite orbits of these elements. 

For the $(A_{3})_{-2}$ case we exploited the fact that the associated Dynkin diagram can be bicoloured. We proved that the Kostant identity (\ref{Kostantid}) also holds for Lorentzian algebras. We also showed that while there are no integer exponents for these algebras, the analogue quantities of them still add up pairwise to $\pi$. These features allowed for a systematic analysis of the respective invariants. We showed that independent invariants at higher degree do not exist, which in turn indicates that the standard approach to utilize these invariants in the construction of integrable systems can not be pursued.

Finally, we used the structures found to formulate models of Calogero type with an in-built infinite symmetry. In their raw form the potential of these models contain as many infinite sums as simple roots of their associated algebras simply constrained by a Diophantine equation. However, using the formulae for the Coxeter orbits we argued that these sums may be reduced. In the case of the invariant affine potential we showed that the remaining single infinite sum may even be computed explicitly. For the models based on the hyperbolic or Lorentzian algebras it remains an open issue how to systematically generate these representatives, whether there are finitely many representatives and whether the sum over the Coxeter orbits can be computed explicitly. We demonstrate, however, that one may perform subsets of the infinite sums obtaining in this manner potentials that are invariant under the action of the Coxeter element. We also stress once more that generating all orbits needed in the potential is different, and more restricted, than generating the entire root space from all orbits. As we have seen for the case of the affine Weyl group when the orbits are generated within the potentials we may use their particular symmetries, such as parity invariance and the fact that we may shift the summation indices. These properties made it possible to construct invariant potentials from a finite number of terms.

Besides the numerous fundamental mathematical issues left open, there are many further interesting questions left for future investigations. The generalisation beyond rational potentials to models of Calogero-Moser-Sutherland type appears to be straightforward. It would also be interesting to consider extended algebraic systems beyond the two examples of $A_2$ and $A_3$ presented here. Of course, ultimately one would also like to understand the quantum versions of the models proposed here.

\medskip

\noindent \textbf{Acknowledgments:}  FC is supported by Fondecyt grant 1211356. AF thanks the Instituto de Ciencias F{\'{\i}}sicas y Matem{\'{a}}ticas of the Universidad Austral de Chile, where part of this work was completed for kind hospitality and Francisco Correa for financial support. OQ thanks the finantial support of Universidad Austral de Chile and Fondecyt grant 1211356 (FC) as well as the Department of Mathematics of City, University of London for kind hospitality in the visit that started this work.

\newif\ifabfull\abfulltrue


\end{document}